\documentclass[a4paper,11pt]{article}
\usepackage{jcappub}
\usepackage{bm}
\usepackage{float}
\usepackage[nooneline]{subfigure}

\usepackage{rotating}
\usepackage{floatpag}
\floatpagestyle{empty}
\usepackage{changepage}
\usepackage{amsmath}
\usepackage{breqn}
\usepackage{placeins}

\title{\boldmath Predicting the intensity mapping signal for multi-$J$ CO lines}

\author[1,2]{Natalie Mashian,}
\author[2]{Amiel Sternberg,}
\author[1]{and Abraham Loeb}
\affiliation[1]{Harvard-Smithsonian Center for Astrophysics, \\60 Garden Street, Cambridge, MA 02138, USA}
\affiliation[2]{The Raymond and Beverly Sackler School of Physics and Astronomy, Tel Aviv University, \\ Tel Aviv 69978, Israel}
\emailAdd{nmashian@physics.harvard.edu}
\emailAdd{amiel@wise.tau.ac.il}
\emailAdd{aloeb@cfa.harvard.edu}

\abstract{We present a novel approach to estimating the intensity mapping signal of any CO rotational line emitted during the Epoch of Reionization (EoR). Our approach is based on large velocity gradient (LVG) modeling, a radiative transfer modeling technique that generates the full CO spectral line energy distribution (SLED) for a specified gas kinetic temperature, volume density, velocity gradient, molecular abundance, and column density. These parameters, which drive the physics of CO transitions and ultimately dictate the shape and amplitude of the CO SLED, can be linked to the global properties of the host galaxy, mainly the star formation rate (SFR) and the SFR surface density. By further employing an empirically derived $SFR-M$ relation for high redshift galaxies, we can express the LVG parameters, and thus the specific intensity of any CO rotational transition, as functions of the host halo mass $M$ and redshift $z$. Integrating over the range of halo masses expected to host CO-luminous galaxies, we predict a mean CO(1-0) brightness temperature ranging from $\sim$ 0.6 $\mu$K at $z$ = 6 to $\sim$ 0.03 $\mu$K at $z$ = 10 with brightness temperature fluctuations of $\Delta$$_{CO}^2$ 
$\sim$ 0.1 and 0.005 $\mu$K respectively, at $k$ = 0.1 Mpc$^{-1}$. In this model, the CO emission signal remains strong for higher rotational levels at $z$ = 6, with $\langle T_{CO} \rangle \sim$ 0.3 and 0.05 $\mu$K for the CO J = 6$\rightarrow$5 and CO J = 10$\rightarrow$9 transitions respectively. Including the effects of CO photodissociation in these molecular clouds, especially at low metallicities, results in the overall reduction in the amplitude of the CO signal, with the low- and high-J lines weakening by 2-20\% and 10-45\%, respectively, over the redshift range 4 $< z <$ 10.}


\begin{document}
\maketitle

\section{Introduction}
Within the last decade, spectral line intensity mapping has been proposed as an additional, complementary probe of the large-scale structure (LSS) of star-forming galaxies during the epoch of reionization (EoR) ~\cite{1999ApJ...512..547S,2008A&A...489..489R,2010JCAP...11..016V}.  Glimpses into this era have been limited to observations of individual massive galaxies and quasars at high redshifts, provided by the Hubble Space Telescope (HST) and ground-based telescopes. While in the future, we anticipate new instruments like Atacama Large Millimeter Array (ALMA) and the James Webb Space Telescope (JWST) providing more detailed views of this ``cosmic dawn", they will be restricted by relatively small fields of view and an inability to observe galaxies that are simply too faint to detect individually. Intensity mapping offers a complementary glimpse of the three-dimensional structure of the high-redshift universe by imaging aggregate line emissions from thousands of unresolved objects and studying the large-scale fluctuations in the given line intensity due to the clustering of unresolved sources.

Intensity mapping can be performed using many different spectral lines, the 21 cm neutral hydrogen line being among the most common, given its unique insight into the evolution of the neutral IGM during the EoR~\cite{1997ApJ...475..429M,2004PhRvL..92u1301L,2004ApJ...608..622Z,2006PhR...433..181F,2010ARA&A..48..127M,2011AAS...21710703P}. In this paper, we focus on the millimeter to far-infrared rotational transitions of carbon monoxide (CO), a molecule that forms primarily in star-forming regions and whose intensity maps promise a wealth of information on the spatial distribution of star formation in the universe~\cite{2014MNRAS.443.3506B}. While CO intensity fluctuations have already been studied, initially as foreground contaminants to cosmic microwave background (CMB) measurements~\cite{2008A&A...489..489R} and then as probes of LSS~\cite{2010JCAP...11..016V,2011ApJ...730L..30C,2011ApJ...728L..46G,2011ApJ...741...70L,2013ApJ...768...15P}, these studies have been limited to the lowest order transitions of the molecule, mainly, CO(1-0) and CO(2-1). These lines are often considered because they are typically among the brightest and have redshifted frequencies that can potentially be observed from the ground. However, with the advent of ALMA\footnote{http://almascience.nrao.edu} and its frequency coverage (84 - 950 GHz), fluctuations in the line emission of many high-J CO transitions will potentially be available and can be measured and translated into a 3D map of the early universe. Having access to multiple CO rotational lines will further facilitate redshift identification and mitigate line confusion, making it possible to statistically isolate the fluctuations from a particular redshift by cross-correlating the emission from different sets of lines~\cite{2010JCAP...11..016V,2013fgu..book.....L}. 

Attempts in the recent literature to obtain a theoretical estimate of the CO emission signal from high redshift galaxies have relied heavily on empirical relations calibrated from local observations and are limited almost exclusively to the $^{12}$CO J=1$\rightarrow$0 transition line. To calculate this mean CO brightness, a simple model is often adopted that connects the strength of the CO emission to the abundance of the dark matter halos that host CO luminous galaxies.~\cite{2010JCAP...11..016V} construct this model by first approximating a galaxy's star formation rate (SFR) as a linear function of the mass of the galaxy's host halo. They then further assume a linear relationship between the line luminosity and SFR and adopt the $L_{CO}$ to SFR ratio from M\,82 to calibrate the proportionality constant.~\cite{2011ApJ...741...70L} embrace a similar approach, adopting the $SFR-M$ relation proposed by~\cite{2010JCAP...11..016V}, but using a set of empirical scaling relations between a galaxy's SFR, far-infrared luminosity, and CO(1-0) luminosity that have been measured for galaxies at $z \lesssim$ 3. Both studies lead to a simple empirical estimate of the CO luminosity that is linear in halo mass ($L_{CO} \propto SFR \propto M$) and that relies on the extrapolation of low-redshift calibrations to higher redshifts, redshifts corresponding to the EoR.~\cite{2011ApJ...728L..46G} arrives at the relation between CO(1-0) luminosity and the halo mass via a different route, making use of the Millennium numerical simulation results of~\cite{2009ApJ...698.1467O}, a study which, although incorporates many physical processes to model the CO emission from high-redshift galaxies, still invokes low-redshift measurements to calibrate the normalization factor of the CO luminosity for a given halo. 

These various models, among others, have led to estimates of the CO power spectra amplitude signal that vary over a range spanning two orders of magnitude, illustrating the lack of theoretical understanding of the physics of CO transitions in a high-redshift context~\cite{2014MNRAS.443.3506B}. In~\cite{2013MNRAS.435.2676M}, this problem is addressed and a computation of CO fluxes is presented within an analytic framework that incorporates both global modes of star formation and the physics of molecular rotational lines in $z \gtrsim$ 6 Lyman-break galaxies. Our paper follows the general direction taken by~\cite{2013MNRAS.435.2676M} and introduces a simpler approach that captures and ties the physics of molecular emission lines to high-redshift ($z \geq$ 4) observations of star-forming galaxies~\cite{2015arXiv150700999M}. Our approach is based on large velocity gradient (LVG) modeling, a radiative transfer modeling technique that generates the full CO spectral line energy distribution (SLED) for a specified set of physical parameters. Typically, LVG modeling is employed to quantitatively analyze an observed set of emission lines and determine the set of parameters that best reproduce the observed SED~\cite{2015ApJ...802...81M}. In this paper, we consider applying the LVG methodology in the reverse direction: given a halo of mass $M$ with CO-emitting molecular clouds characterized by a kinetic temperature $T_{kin}$, velocity gradient $dv/dr$, gas volume density $n$, molecular abundance $\chi$, and column density $N$, we will derive the full CO SED and compute the mean surface brightness of any CO rotational line emitted by halos with that mass at any given redshift. 

Our paper is organized as follows. In Section 2, we introduce our LVG model for the specific intensity of CO emission and outline the formalism that relates the set of LVG parameters driving the physics of CO transitions in molecular clouds to the global properties of the host galaxy, mainly, the SFR. Given the empirically determined high-redshift $SFR-M$ relation~\cite{2015arXiv150700999M}, these parameters, and thus, the specific CO intensity, can ultimately be expressed as functions of the host halo mass $M$ and redshift $z$. In Section 3 we compute the spatially averaged CO surface brightness from star-forming galaxies, as well as the power spectrum of spatial fluctuations in the CO emission at any given redshift, with a focus on redshifts corresponding to the EoR. We conclude in Section 4 with a summary of our results and a brief comparison with other related calculations of the CO intensity mapping signal. Throughout we consider a $\Lambda$CDM cosmology parametrized by $n_s$ = 1, $\sigma_8$ = 0.8, $\delta_c$ = 1.69, $\Omega_m$ = 0.31, $\Omega_\Lambda$ = 0.69, $\Omega_b$ = 0.05, and $h$ = 0.7, consistent with the latest measurements from Planck~\cite{2015arXiv150201589P}.

\section{Modeling the CO Emission}

\subsection{CO Brightness Temperature}
To calculate the average CO brightness temperature, we follow the formalism presented by~\cite{2011ApJ...741...70L} and consider the specific intensity of a CO line observed at frequency $\nu_{obs}$ at redshift $z$ = 0,
\begin{equation}
I(\nu_{obs})=\frac{c}{4\pi}\int_0^\infty dz'\,\,\frac{\epsilon\left[\nu_{obs}(1+z')\right]}{H(z')(1+z')^4}
\end{equation}
as determined by solving the cosmological radiative transfer equation, where $H(z)$ is the Hubble parameter and $\epsilon\left[\nu_{obs}(1+z')\right]$ is the proper volume emissivity of the given line. Since CO is emitted from within halos hosting star-forming galaxies, we take the CO luminosity, $L_{CO}$, to be some function of the halo mass $M$ and redshift, and assume the profile of each CO line is a delta function in frequency,
\begin{equation}
L_{CO}=L(M,z)\delta_D(\nu-\nu_J)
\end{equation}
where $\nu_J$ is the rest frame frequency of the transition of interest. If we then further assume that at any given time, a fraction $f_{duty}$ of halos with mass larger than $M_{min,CO}$ actively emit CO lines, then for a given halo mass function $dn/dM$, the volume emissivity is,
\begin{equation}
\epsilon(\nu,z)=\delta_D(\nu-\nu_J)(1+z)^3f_{duty}\int_{M_{min,CO}}^\infty\hspace{-0.7cm} dM \,\,\frac{dn}{dM}(M,z)L(M,z)
\end{equation}

The specific intensity of a line with rest frame frequency $\nu_J$, emitted by gas at redshift $z_J$ thus simplifies to
\begin{equation}
I_{\nu_{obs}}=\frac{c}{4\pi}\frac{1}{\nu_J H(z_J)}f_{duty}\int_{M_{min,CO}}^\infty\hspace{-0.7cm} dM\,\,\frac{dn}{dM}(M,z_J)L(M,z_J)
\end{equation}
or, written as the brightness temperature,
\begin{equation}
\langle T_{CO} \rangle = \frac{c^3}{8\pi}\frac{(1+z_J)^2}{k_B \nu_J^3 H(z_J)}f_{duty}\int_{M_{min,CO}}^\infty\hspace{-0.7cm} dM\,\,\frac{dn}{dM}(M,z_J)L(M,z_J) 
\end{equation}
where $k_{B}$ is the Boltzmann constant.

To determine $L(M,z_J)$, the specific luminosity of a given CO line, we employ large velocity gradient (LVG) modeling, a method of radiative transfer in which the excitation and opacity of CO lines are determined by the kinetic temperature $T_{kin}$, velocity gradient $dv/dr$, gas density $n$, CO-to-H$_2$ abundance ratio $\chi_{CO}$, and the CO column density of the emitting source. We adopt the escape probability formalism~\cite{1970MNRAS.149..111C,1974ApJ...189..441G} derived for a spherical cloud undergoing uniform collapse where
\begin{equation}
\beta_J=\frac{1-e^{\tau_{J}}}{\tau_{J}}
\end{equation}
is the probability, for a line optical depth $\tau_J$, that a photon emitted in the transition $J\rightarrow J-1$ escapes the cloud (see~\cite{2013MNRAS.435.2407M} for a more detailed presentation of the LVG formalism.) Assuming that each emitting source consists of a large number of these unresolved homogeneous collapsing clouds, the corresponding emergent intensity of an emission line integrated along a line of sight can be expressed as
\begin{equation}
I_J=\frac{h\nu_J}{4\pi} A_J x_J\beta_J(\tau_J)N_{CO}
\end{equation}
where $x_J$ is the population fraction in the $J^{th}$ level, $h\nu_J$ is the transition energy, $A_J$ is the Einstein radiative coefficient, and $N_{CO}$ is the beam-averaged CO column density.  
The LVG-modeled specific luminosity of a line emitted by a host halo with disk radius $R_{d}$,  therefore takes the form
\begin{equation}
L_{CO}(M,z_J) = 4\pi^2R_{d}^2I_{J,LVG}=\pi h\nu_J R_{d}^2 A_J x_J\beta_J(\tau_J)N_{CO} \,\,.
\end{equation}
We set
\begin{eqnarray}
R_d(M,z) &=& \frac{\lambda}{\sqrt{2}}\frac{j_d}{m_d} r_{vir}\nonumber\\
&=& \frac{\lambda}{\sqrt{2}}\frac{j_d}{m_d}\times 1.5 \left[\frac{\Omega_m}{\Omega_m(z)}\frac{\Delta_c}{18\pi^2}\right]^{-1/3}\left(\frac{M}{10^8 M_\odot}\right)^{1/3}\left(\frac{1+z}{10}\right)^{-1}  \textrm{kpc}
\end{eqnarray}
where $\Delta_c = 18\pi^2+82d-39d^2$, $d = \Omega_m(z)-1$, and $\Omega_m(z) = \Omega_m(1+z)^3/(\Omega_m(1+z)^3+\Omega_\Lambda)$~\cite{2013fgu..book.....L}. We assume that the specific angular momentum of the material that forms the disk is the same as that of the halo, i.e. $j_d/m_d = 1$, and adopt a spin parameter of $\lambda\approx$ 0.05, corresponding to an isolated exponential disk~\cite{1998MNRAS.295..319M}. 

Since the excitation state and optical depth of a given line, $x_J$ and $\tau_J$  respectively, are determined by the set of physical parameters \{$T_{kin}$, $dv/dr$, $n$, $\chi_{CO}$\} that characterize the emitting molecular clouds, the task remains to express these parameters as functions of $M$ and $z$, global properties of the host halo.  

\subsection{The Star Formation Model}

\FloatBarrier
\begin{figure*}[t!]
\begin{minipage}{1\linewidth}
\hspace{-1.7cm}\includegraphics[width=575pt,height=325pt]{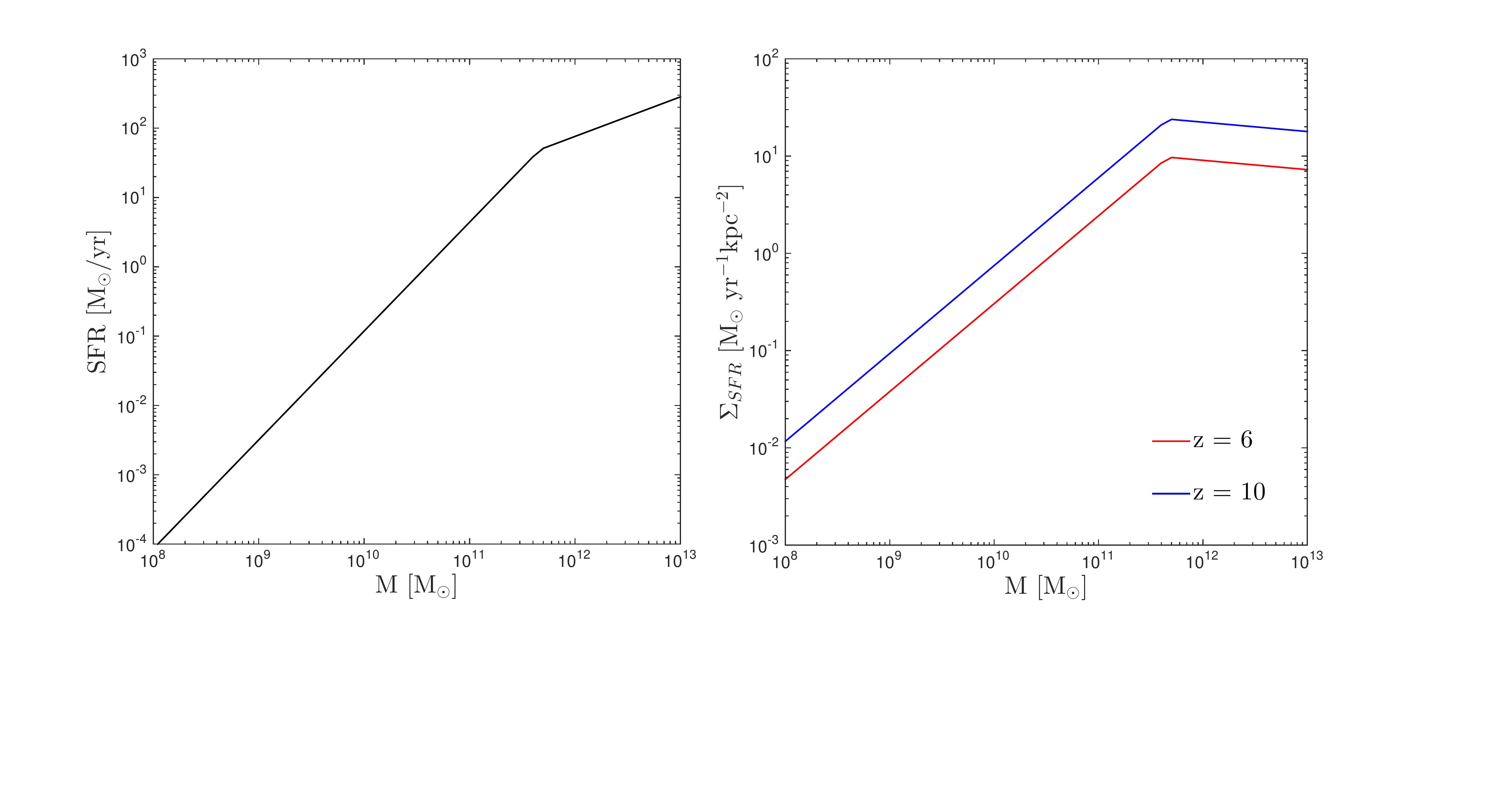}
\vspace{-3cm}
\caption{\textit{Left panel:} The mean $SFR-M$ relation in the high redshift universe, i.e. $z \gtrsim$ 4, derived empirically via abundance matching and fitted by the double power law given in eq. (2.11) where \{$a_1, a_2, b_1, b_2$\} = \{2.4$\times$10$^{-17}$, 1.1$\times$10$^{-5}$, 1.6, 0.6\} for $f_{UV}$ = 1 . \textit{Right panel:} The corresponding SFR surface density, $\Sigma_{SFR}(M,z)$, derived by dividing the SFR by the halo disk area, $\pi R_d(M,z)^2$, at redshifts $z$ = 6 (red) and $z$ = 10 (blue). For more details and comparison to data, see~\cite{2015arXiv150700999M}.}
\end{minipage}
\end{figure*}

As will be physically motivated below, the LVG parameters that dictate the shape of CO SLEDs in galaxies are well correlated with the galaxy's global star formation rate surface density. The first ingredient of our model is therefore a $SFR-M$ relation that connects the SFR and host halo mass at the high redshifts we are concerned with in this paper. In~\cite{2015arXiv150700999M}, such an empirical relation is derived by mapping the shape of the observed ultraviolet luminosity functions (UV LFs) at $z \sim$ 4-8 to that of the halo mass function at the respective redshifts. In this abundance-matching method, each dark-matter halo is assumed to host a single galaxy and the number of galaxies with star formation rates greater than $SFR$ are equated to the number of halos with mass greater than $M$,
\begin{equation}
f_{UV}\int_M^\infty \,\,dM\,\,\frac{dn}{dM}(M,z) = \int_{SFR}^\infty \,\, dSFR\,\,\phi(SFR,z)
\end{equation}
where $\phi(SFR,z)$ is the observed, dust-corrected UV LF at redshift $z$ and $f_{UV}$ is the starburst duty cycle, i.e.\,the fraction of halos with galaxies emitting UV luminosity at any given time.~\cite{2015arXiv150700999M} find that the $SFR-M$ scaling law remains roughly constant across this redshift range and thus, an average relation can be obtained and applied to even higher redshifts, $z \gtrsim$ 8, where it faithfully reproduces the observed $z \sim$ 9 and 10 LFs. This mean scaling law, $SFR_{av}(M)$, is fairly well parameterized by a double power law of the form,
\begin{equation}
SFR_{av}(M) =  
\begin{cases}
   a_1M^{b_1}\,, & M\leq M_c\\
    a_2M^{b_2}\,,    & M\geq M_c
\end{cases}
\end{equation}
with a turnover at a characteristic halo mass $M_c \approx$ 10$^{11.6}$ M$_\odot$. Fitting the average relations in the observed SFR range $\simeq$ 0.1 - 500 M$_\odot$/yr, we obtain \{$a_1, a_2, b_1, b_2$\} = \{2.4$\times$10$^{-17}$, 1.1$\times$10$^{-5}$, 1.6, 0.6\} for $f_{UV}$ = 1. We find it reasonable to assume a UV duty cycle of unity throughout our calculations given that the time between mergers grows shorter than the Hubble time at the high redshifts we are considering and the fact that typical hydrodynamical simulations, where star-formation is driven not just by mergers, find a star-forming galaxy in effectively every halo at these redshifts~\cite{2010MNRAS.406.2267F}.

 The corresponding mean SFR surface density, $\Sigma_{SFR}$ (units M$_\odot$ yr$^{-1}$kpc$^{-2}$), is computed by dividing the SFR from eq. (2.11) by the area of the active star-forming halo disk with radius given by eq. (2.9). Figure 1 depicts the average $SFR-M$ relation and the resulting SFR surface densities at redshifts $z$ = 6 and 10.

\subsection{Theoretical Models for LVG Parameters}
The next key step in our LVG-motivated approach to predicting the high redshift CO emission signal is to model the LVG parameters dictating the shape of the CO SLED as functions of the global properties of the host halo. As pointed out in~\cite{2014MNRAS.442.1411N}, quantities, such as the gas temperature and density, which characterize the CO-emitting molecular interstellar medium (ISM) are well-correlated with the star formation rate surface densities of galaxies. Qualitatively, this makes sense since regions of high SFR density typically arise from denser gas concentrations and have large UV radiation fields, with increased efficiency for thermal coupling between gas and dust. In the following sections, we will outline how each LVG parameter can be expressed in terms of the SFR surface density, $\Sigma_{SFR}$, and thus ultimately as a function of just the halo mass and redshift. 

\subsubsection{Gas Kinetic Temperature}
To determine the effective kinetic temperature of the CO-emitting molecular gas, we assume that the gas and dust in the star-forming disk are thermally well-coupled, i.e. $T_{gas} \approx T_{dust}$. The temperature of a dust grain is set by the balance of radiative heating and cooling processes taking place in molecular clouds. The rate of heating due to absorption of optical or UV radiation and the incident cosmic microwave background (CMB) can be written as
\begin{equation}
\left(\frac{dE}{dt}\right)_{abs}=\pi a^2\left[\int_0^\infty d\nu \,\,Q_{abs,UV}(\nu)F_{\nu}(\nu)\,+\,\sigma_{SB}T_{CMB}^4\right]
\end{equation}
where $\sigma_{SB}$ is the Stefan-Boltzmann constant, $a$ is the grain radius, $Q_{abs,UV}(\nu)$ is the emissivity in the UV-optical regime, which we set equal to unity~\cite{2007A&A...462...81C}, $T_{CMB}$ = 2.73(1+$z$) is the CMB temperature and $F_\nu(\nu)$ is the flux of energy radiated by the central starburst in the disk. Assuming optically thick conditions requires this starlight energy to be totally reemitted in the infrared; the integral on the right-hand side can thus be expressed as $L_{IR}/4\pi R_d^2$ and, adopting the conversion between $L_{IR}$ and $SFR$ presented in~\cite{1998ApJ...498..541K},
\begin{equation}
\left(\frac{L_{IR}}{\text{erg s$^{-1}$}}\right)=2.2\times10^{43}\left(\frac{SFR}{\text{M$_\odot$ yr$^{-1}$}}\right) \,,
\end{equation}
eq. (2.12) takes the final form:
\begin{equation}
\left(\frac{dE}{dt}\right)_{abs}=\pi a^2 \left[\frac{2.2\times10^{43}}{4\pi}\frac{\Sigma_{SFR}}{\text{(M$_\odot\, $yr$^{-1}$kpc$^{-2}$)}}\,+\,\sigma_{SB}(2.73(1+z))^4\right] \,\,.
\end{equation}

The rate of cooling of dust grains by infrared emission is given by
\begin{equation}
\left(\frac{dE}{dt}\right)_{em}=4\pi a^2\int_0^\infty d\nu\,\,Q_{abs,IR}(\nu)\pi B_\nu(\nu,T_d)
\end{equation}
where $Q_{abs,IR}(\nu) \simeq Q_{abs}(\nu_0)(\nu/\nu_0)^\beta$ is the emissivity in the infrared regime, $\nu_0$ is the reference frequency, and $\beta$ is the dust emissivity index. 
Substituting in for $Q_{abs,IR}(\nu)$ and $\pi B_\nu(\nu,T_d)$ (the flux emitted by a black-body), the integral simplifies to
\begin{equation}
\left(\frac{dE}{dt}\right)_{em}=\pi a^2\left[\frac{60\sigma_{SB}}{\pi^4}\left(\frac{k_B}{h}\right)^\beta\frac{Q_{abs}(\nu_0)}{\nu_0^\beta}\Gamma(\beta+4)\zeta(\beta+4)T_d^{\beta+4}\right] \,\,.
\end{equation}

\noindent At thermal equilibrium, $(dE/dt)_{abs}=(dE/dt)_{em}$, and the steady-state grain temperature in units of kelvin is
\begin{dmath}
T_d(\Sigma_{SFR}(M,z),z) = \left(\frac{\pi^4}{60}\left(\frac{h}{k_B}\right)^\beta\frac{\nu_0^\beta}{Q_{abs}(\nu_0)}\frac{1}{\Gamma(\beta+4)\zeta(\beta+4)}\left[\frac{2.3\times10^{-3}}{4\pi\sigma_{SB}}\frac{\Sigma_{SFR}(M,z)}{\text{(M$_\odot$\,yr$^{-1}$kpc$^{-2}$)}}+(2.73(1+z))^4\right]\right)^{1/(\beta+4)}\hspace{-1cm} . 
\end{dmath}
In the analysis below, we take the fiducial value of $\beta$ = 1.3, consistent with the mean emissivity index derived from the SCUBA Local Universe Galaxy Survey~\cite{2000MNRAS.315..115D}, and the standard value $Q_{abs}$(125 $\mu$m) = 7.5$\times$10$^{-4}$~\cite{1983QJRAS..24..267H,2004A&A...425..109A,2007A&A...462...81C}. A plot of the kinetic temperature as a function of halo mass at different redshifts can be found in the left panel of figure 2.

\subsubsection{Cloud volume density}
Another key parameter in determining the shape of the CO SLED is $n$, the cloud volume density (cm$^{-3}$) of the dominant collision partner for CO rotational excitation. Since we are assuming CO-emitting molecular clouds, $n$ in this case is the cloud volume density of H$_2$, given by
\begin{equation}
n_{H_2}=\frac{3\Sigma_{cl}}{4\mu m_{H_2}r_{cl}}
\end{equation}
where $\mu$ = 1.36 takes into account the helium contribution to the molecular weight (assuming cold, neutral gas), $m_{H_2}$ = 3.34$\times$10$^{-27}$ kg, and $r_{cl}$ is the cloud radius assuming a uniform sphere. At these high redshifts where the ISM is dominated by molecular gas, the surface density of H$_2$ in a given molecular cloud, $\Sigma_{cl}$, can be related to the beam-averaged gas surface density in the galactic disk, $\Sigma_{gas}$.

Empirical studies have found that a correlation exists between the surface density of molecular gas and the surface density of the SFR, a discovery that is consistent with observations that stars form predominantly in the molecular component of the ISM. The Kennicutt-Schmidt (KS) relation~\cite{1959ApJ...129..243S,1989ApJ...344..685K} formulates this correlation in terms of a power-law, $\langle\Sigma_{SFR}\rangle \propto \langle\Sigma_{gas}\rangle^N$, where estimates of the index $N$ range from super-linear~\cite{1989ApJ...344..685K,1998ApJ...498..541K,2007ApJ...671..303B,2011ApJ...735...63L,2013ApJ...772L..13M}, to linear~\cite{2008AJ....136.2846B,2008AJ....136.2782L,2013AJ....146...19L}, to sublinear ~\cite{2014MNRAS.437L..61S}. 
Given our paper's focus on high redshift star-forming sources with CO-emitting molecular clouds, we deemed it most appropriate to adopt the KS relation presented in~\cite{2010MNRAS.407.2091G},
\begin{equation}
\Sigma_{SFR} =(3.3\pm0.6)\times10^{-4} \left(\frac{\Sigma_{gas}}{1 \text{ M$_\odot$ pc$^{-2}$}}\right)^{1.2\pm0.1} \,\,\,\text{M$_\odot$ yr$^{-1}$ kpc$^{-2}$} \,\, ,
\end{equation}
a relation that was derived from data sets of CO molecular emission in $z \sim$ 1-3 normal star-forming galaxies and restricted to the regime where molecular gas dominates the ISM at these redshifts, i.e. $\Sigma_{gas} \gtrsim$ 3 M$_\odot$ pc$^{-2}$. Since no evidence of any redshift-dependence of this relation has been found thus far, applying eq. (2.19) at the higher redshifts we consider in this paper, $z \geq$ 4, is a reasonable extrapolation.

We therefore set $\Sigma_{cl}$ equal to $\Sigma_{gas}$, as defined by eq. (2.19), except in cases where $\Sigma_{gas}$ drops below the threshold surface density for which the cloud is predominantly molecular. This ``star-formation" threshold, defined by a molecular gas fraction of $f_{H_2}$ = 0.5, is derived in~\cite{2014ApJ...790...10S} for a plane-parallel slab (including H$_2$ dust) as a function of metallicity $Z'$,
\begin{equation}
\Sigma_{gas,*}(Z')=\frac{2m}{\sigma_g(Z')}\left(1.6\ln{\left[\frac{\alpha G(Z')}{3.2}+1\right]}\right)
\end{equation}
where $m$ = 2.34$\times$10$^{-27}$ kg is the mean particle mass per hydrogen nucleus, $\sigma_g(Z') = 1.9\times10^{-21}Z'$ cm$^{-2}$ is the dust-grain Lyman-Werner-photon absorption cross section per hydrogen nucleon, and $\alpha G$ is the dimensionless parameter that defines the LW-band optical depth in the cloud due to HI dust,
\begin{equation}
\alpha G(Z') = \left(\frac{1+3.1Z^{'0.365}}{4.1}\right)\frac{6.78}{1+\sqrt{2.64Z'}}\,\,.
\end{equation}

\FloatBarrier
\begin{figure*}[t!]
\begin{minipage}{1\linewidth}
\hspace{-2cm}\includegraphics[width=575pt,height=325pt]{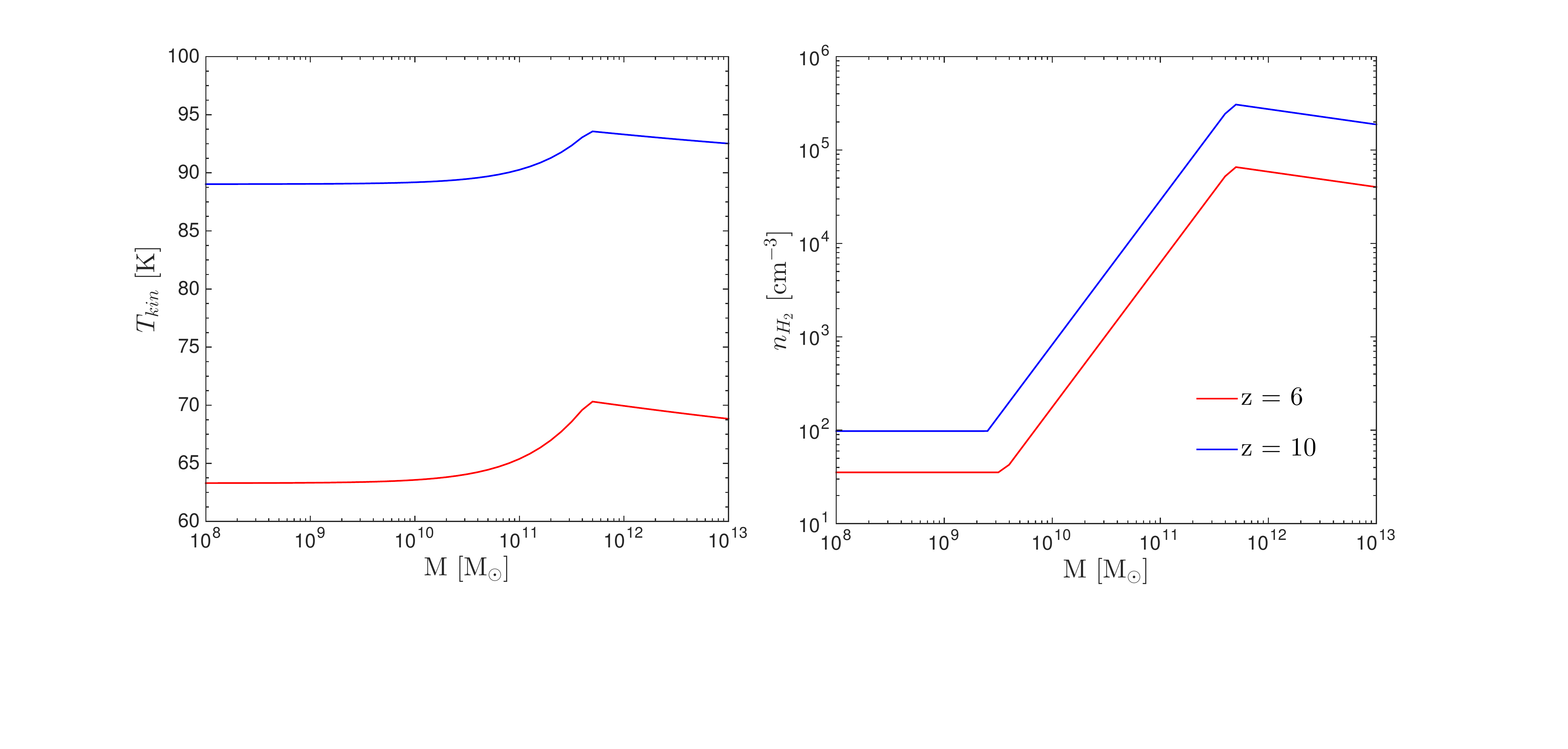}\\
\vspace{-3cm}\caption{\textit{Left panel:} Gas kinetic temperature assuming the fiducial values $\beta$ = 1.3 and $Q_{abs}$(125 $\mu$m) = 7.5$\times$10$^{-4}$ in eq. (2.17). \textit{Right panel:} Cloud volume density of molecular hydrogen as defined by equations (2.18)-(2.27) plotted at redshifts $z$ = 6 (red) and $z$ = 10 (blue).}
\end{minipage}
\end{figure*}

We adopt the fundamental metallicity relation (FMR), a tight relation between the gas-phase metallicity $Z'$, stellar mass $M_*$, and SFR, to ultimately express $Z'$ as a function solely of the halo mass and redshift. The FMR was initially observed and formulated in~\cite{2010MNRAS.408.2115M} for local galaxies in the mass range 9.2 $\leq \log{M_*/M_\odot} \leq$ 11.4. Since then, the FMR has been confirmed to hold for star-forming galaxies at redshifts as high as $z \sim$ 3~\cite{2013ApJ...772..141B}, and to extend smoothly at lower masses~\cite{2011MNRAS.414.1263M}, taking the final form,
\begin{equation}
12+\log{(O/H)}=
\begin{cases}
8.90+0.37m-0.14s-0.19m^2+0.12ms-0.054s^2\, & \text{for}\,\,\,\, \mu_{0.32}\geq9.5\\
8.93+0.51(\mu_{0.32}-10)\, & \text{for}\,\,\,\,  \mu_{0.32} < 9.5
\end{cases}
\end{equation}
where $\mu_\alpha = \log{(M_*)} - \alpha\log{(SFR)}$, $m = \log{(M_*)}-10$, and $s=\log{(SFR)}$.

To further parametrize the metallicity as a function of the halo mass and redshift, we rely on observations and models that support the conclusion that the SFR in galaxies at redshifts $z \gtrsim$ 4 scales nearly linearly with increasing stellar mass and does not vary by more than a factor of order 2~\cite{2015ApJ...799..183S}. This behavior is consistent with a crude estimation of the stellar mass of a galaxy with a star formation rate $SFR$:
\begin{equation}
M_* \sim \int_0^{t_H(z)}dt\,SFR(t)\sim SFR(z)\times t_H(z)
\end{equation}
where $t_H(z)$ is the age of the universe at a given redshift $z$. Since each halo of interest formed at some fraction of the age of the universe, the right-hand side should be multiplied by a factor $\lesssim$ 1. We therefore calibrate the above expression using the SFR-$M_*$ best-fit parameters presented in~\cite{2015ApJ...799..183S} for $z \sim$ 4 - 6 and obtain the following relation
\begin{equation}
M_*(M,z)=(0.28\pm0.02)\,SFR_{av}(M)\,t_H(z) 
\end{equation}
 where $SFR_{av}(M)$ is the average SFR for a halo of mass $M$. We assume this relation continues to apply at redshifts $z >$ 6 in the following calculations.

Armed with the parametrization of $SFR$ introduced in \S2.2, and thus a metallicity $Z$ expressed solely as a function of halo mass and redshift, we can now return to our model for the cloud surface density, $\Sigma_{cl}$. To ensure the molecular state of each individual cloud, i.e. $f_{H_2} \geq$ 0.5, we set $\Sigma_{cl}$ equal to the beam-averaged gas surface density $\Sigma_{gas}(M,z)$ (derived by inverting eq. (2.19)) for all halo masses $M > \tilde{M}$ and floor it to the value $\Sigma_{gas,*}(\tilde{M},z,f_{H_2}=0.5)$ for all $M <\tilde{M}$,
\begin{equation}
\Sigma_{cl}(M,z)=
\begin{cases}
\Sigma_{gas}(M,z)\, & \text{if}\,\,\,\, M>\tilde{M}\\
\Sigma_{gas,*}(\tilde{M},z,f_{H_2}=0.5) \, &\text{otherwise}
\end{cases}
\end{equation}
where $\tilde{M}$ is the halo mass at which $\Sigma_{gas}(z)$ drops below $\Sigma_{gas,*}(f_{H_2}=0.5,z)$ at a given redshift $z$.

The other variable that appears in our definition of the H$_2$ volume density is $r_{cl}$, the molecular cloud radius. Assuming that the molecular gas within the clump is in hydrostatic equilibrium, the radius of the cloud can be related to its surface mass density through the relation,
\begin{equation}
r_{cl}= \frac{c_{s,eff}^2}{\pi G \Sigma_{cl}}
\end{equation}
where $G$ is the gravitational constant and $c_{s,eff}$ is the effective sound speed in the gas (taking into account turbulence, $c_{s,eff}^2$ = $c_s^2 + \sigma_{turbulence}^2$), which we set to 10 km/s. The final expression for the molecular cloud volume density then simplifies to
\begin{equation}
n_{H_2}(M,z)=\frac{3\pi G}{4\mu m_{H_2}c_{s,eff}^2}\Sigma_{cl}^2(M,z) \, ,
\end{equation}
a plot of which can be found in the right panel of figure 2. The steep decline in number density at low halo masses, as depicted in figure 2, mirrors the steeply declining $SFR-M$ relation at the low-mass end (see left panel in figure 1); as the star-formation rate diminishes by several orders of magnitude with decreasing halo mass, the SFR surface density drops accordingly and ultimately translates into reduced gas column and number densities at these low masses.

\subsubsection{Velocity Gradient}
For the sake of simplicity, we assume self-gravitating, virialized molecular clouds, in which case, the velocity gradient $dv/dr$ and cloud volume density $n_{H_2}$ are related in the following way~\cite{2001ApJ...557..736G}
\begin{equation}
\frac{dv}{dr}\simeq3.1\sqrt{\frac{n_{H_2}}{10^4\,\,\text{cm$^{-3}$}}}\,\,\,\,\,\text{km\,s$^{-1}$pc$^{-1}$}
\end{equation}
where $n_{H_2}$ is defined in eq. (2.27). A plot of $dv/dr$ as a function of halo mass at different redshifts can be found in the left panel of figure 3.

\FloatBarrier
\begin{figure*}[t!]
\begin{minipage}{1\linewidth}
\hspace{-2cm}\includegraphics[width=575pt,height=325pt]{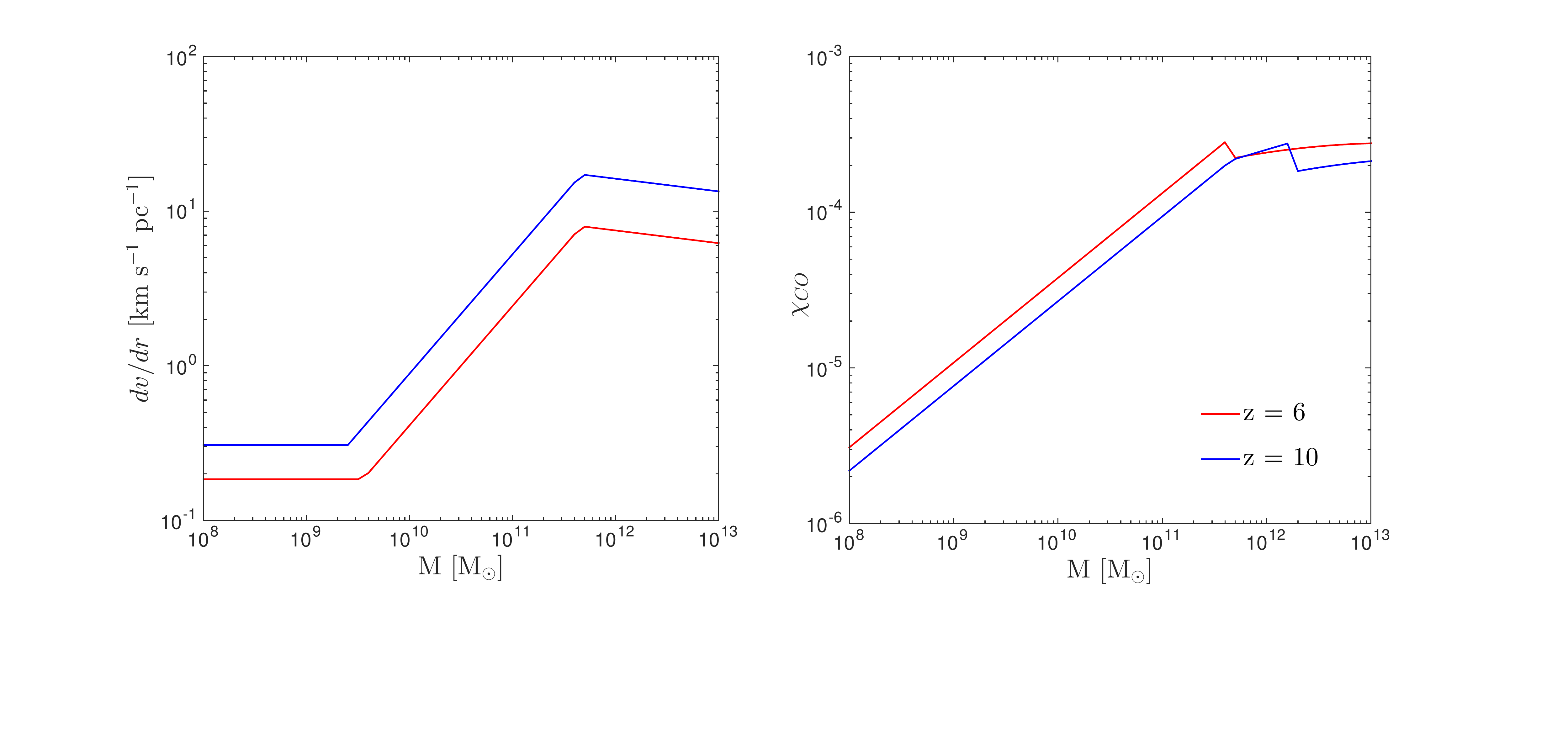}
\vspace{-3cm}\caption{\textit{Left panel:} Velocity gradient $dv/dr$ as a function of halo mass $M$ assuming self-gravitating, virialized clouds as defined by eq. (2.28). \textit{Right panel:} CO-to-H$_2$ abundance ratio via eq. (2.29), plotted at redshifts $z$ = 6 (red) and $z$ = 10 (blue).}
\end{minipage}
\end{figure*}

\subsubsection{CO-to-H$_2$ Abundance Ratio}
Studies have shown that at high metallicities, i.e. $Z' \gtrsim$ 10$^{-2}$, the dominant metal-bearing molecule in the ISM is CO. Furthermore,~\cite{2015MNRAS.450.4424B} find that even at low metallicities, 30-100\% of the available carbon is always locked in CO if the CO is shielded, provided that the hydrogen gas is in molecular form. In this limit, the relative CO-to-H$_2$ abundance varies approximately linearly with metallicity~\cite{2015MNRAS.450.4424B}, and assuming most of the carbon is in fact locked up in CO, $\chi_{CO}$ is given by,
\begin{equation}
\chi_{CO}(Z)\simeq3\times10^{-4}Z'
\end{equation}
where an expression for the relevant metallicity $Z'$ can be found in eq. (2.22). A plot of $\chi_{CO}$ is shown in the right panel of figure 3.

\subsubsection{CO Column Density, including photodissociation}

\FloatBarrier
\begin{figure*}[t!]
\begin{minipage}{1\linewidth}
\hspace{1.5cm}\includegraphics[width=350pt,height=270pt]{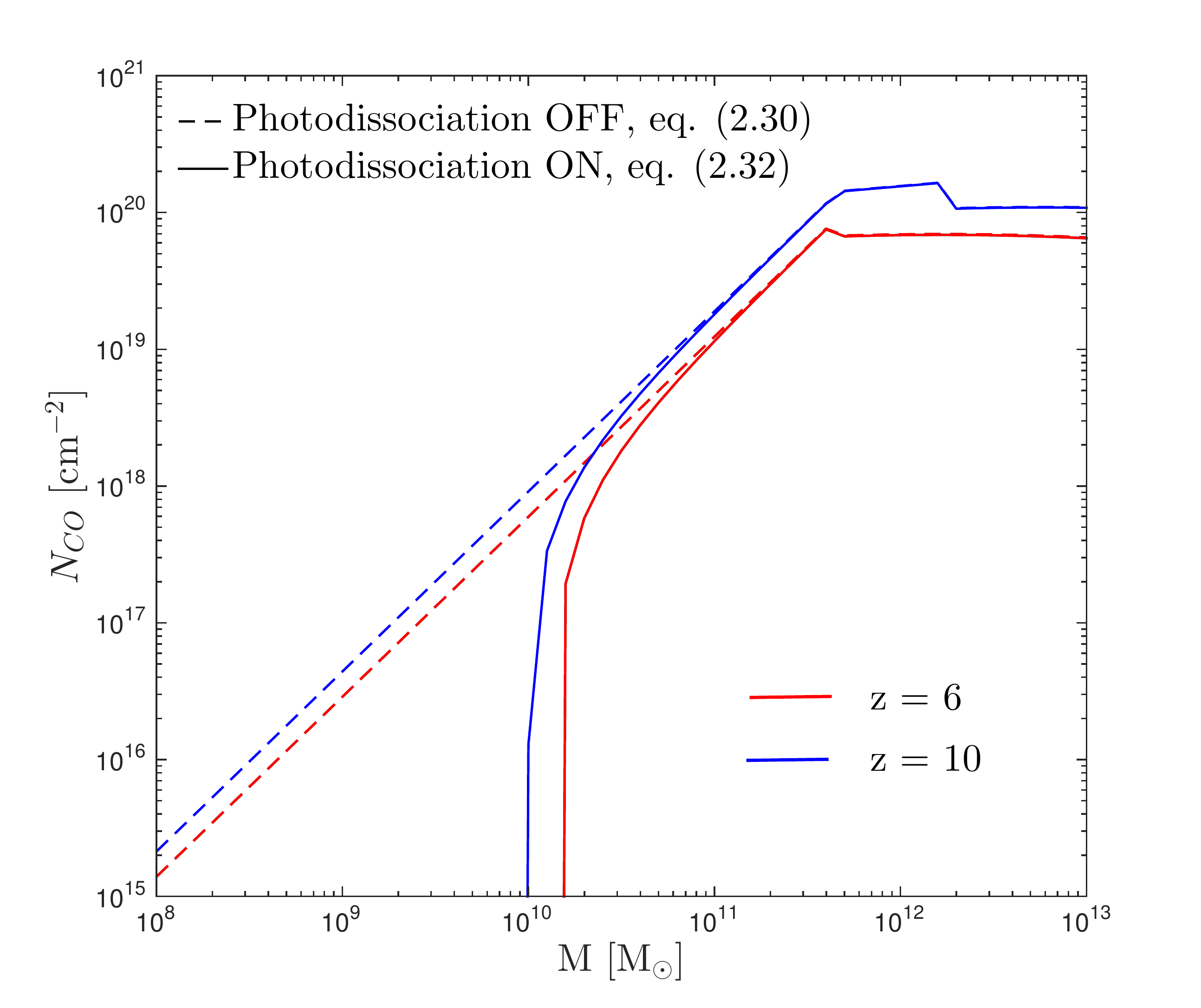}
\vspace{-.5cm}\caption{Beam-averaged CO column density derived at redshifts z=6 (red) and z=10 (blue)  under conditions where CO photodissociation is accounted for (solid curves, eq. (2.30)) and neglected (dashed curves, eq. (2.32)). In the case where the effects of CO photodissociation are included, the CO column density drops to zero for halos smaller than $M\lesssim$ 10$^{10}$ M$_\odot$, indicating that the CO in the gas has fully disassociated.}
\end{minipage}
\end{figure*}

The CO column density, which sets the overall scaling and amplitude of the CO SLED, can be expressed most simply as the product of the CO-to-H$_2$ abundance ratio and the beam-averaged molecular column density,
\begin{equation}
N_{CO}=\chi_{CO}N_{H_2}
\end{equation}
 where $N_{H_2}$ is a power-law function of the SFR surface density, derived by inverting the KS relation in eq. (2.19),
\begin{equation}
N_{H_2}(\Sigma_{SFR}(M,z))=\frac{2\times10^{-4}}{\mu m_{H_2}}\left(\frac{\Sigma_{SFR}}{1 \text{ M$_\odot$ yr$^{-1}$ kpc$^{-2}$}}\right)^{1/1.2}  \,\,\,\text{cm$^{-2}$} \,\,\,\,\, .
\end{equation}

The above expression for $N_{CO}$ does not account for conditions under which CO has photodissociated into C and C$^+$ while the gas continues to remain molecular due to either H$_2$ self-shielding or dust-shielding. Such conditions, which may exist on the surfaces of molecular clouds or the clumps contained within such clouds, ultimately result in a fraction of H$_2$ gas that is ``dark" in CO transitions. Observations indicate that the column density of this ``dark gas" can be as high as 30\% ($\sim$ 3$\times$10$^{21}$ cm$^{-2}$) of the total molecular column density in the local Galaxy ($\sim$10$^{22}$ cm$^{-2}$)~\cite{2010ApJ...716.1191W}. A first-order approximation of the CO column density that accounts for the effects of CO photodissociation is then 
\begin{equation}
N_{CO}=\chi_{CO}\left(N_{H_2} -\frac{3\times10^{21}}{Z'}\right) \,\,\,\text{cm$^{-2}$} 
\end{equation}
where, as before, $Z'$ is the metallicity in solar units. The second term in the above expression is essentially the H$_2$ column required to enable CO dust-shielding under the assumption that the threshold for CO survival is set by a universal dust opacity. This approximation captures the effects of decreasing metallicity on the abundance of CO, reducing the column density as the number of shielding dust grains grows sparse. In the limiting case where $Z'$ drops so low that there is not enough dust to effectively shield CO, (corresponding to the parenthesized portion of eq. (2.32) growing negative), the CO is fully dissociated and $N_{CO}$ is set to zero. We note that eq. (2.32) is only a first-order approximation and that the constant, 3$\times$10$^{21}$, can be altered due to variations in UV field intensity or gas clumping factors. The beam-averaged CO column density at different redshifts is plotted in figure 4, both for the case where CO photodissociation is accounted for (solid curves) and neglected (dashed curves).

\subsection{Model CO SLEDs}
Equipped with analytic expressions for the LVG parameters that dictate the shape and magnitude of the CO SLED, we can now compute the intensity of each CO line and the resulting SED generated by the molecular clouds in a halo with mass $M$ at redshift $z$. To carry out the computations, we use the Mark \& Sternberg LVG radiative transfer code described in~\cite{2012A&A...537A.133D}, with CO-H$_2$ collisional coefficients taken from~\cite{2010ApJ...718.1062Y} and energy levels, line frequencies, and Einstein $A$ coefficients taken from the Cologne Database for Molecular Spectroscopy (CDMS). For a given set of parameters, \{$T_{kin}$, $n_{H_2}$, $dv/dr$, $\chi_{CO}$, $N_{H_2}$\}, the code determines the level populations by iteratively solving the equations of statistical equilibrium which balance radiative absorptions, stimulated emission, spontaneous emission, and collisions with H$_2$ using the escape probability formalism discussed in \S2.1. Once the level populations are computed, the full CO rotational ladder and line intensities follow from eq. (2.7). 


Figure 5 shows the CO SLEDs generated by halos at redshift $z$ = 10 with masses in the range $M$ = 10$^8$-10$^{13}$ M$_\odot$ when the effects of CO photodisassociation are excluded. In the case where photodissociation is considered, the line intensities emitted by low-mass halos with $M <$ 10$^{10}$ M$_\odot$ (red and orange curves) entirely disappear, reflecting the full dissociation of CO in these molecular clouds where dust-shielding has grown inefficient. However, the other curves, corresponding to line emission from higher mass halos that have been normalized to the ground state, remain unchanged. This is due to the fact that ``turning on" photodissociation merely reduces the beam-averaged CO column density according to eq. (2.32); since $N_{CO}$ controls the overall amplitude of the SLED (and not the shape), adjusting this quantity simply amplifies or reduces the intensity of all the CO lines by the same amount, leaving the ratio between lines unchanged. 

 As expected, we find that as the physical conditions in the emitting molecular clouds grow more extreme, the CO rotational levels become increasingly populated and the SLED rises accordingly. Consequently, the line intensities not only grow in magnitude, but the peak of the CO SLED also shifts to higher $J$ values, reflecting the excitation of the more energetic states of the molecule. 

\begin{figure*}[h!]
\begin{minipage}{1\linewidth}
\hspace{-1.5cm} \includegraphics[width=570pt,height=270pt]{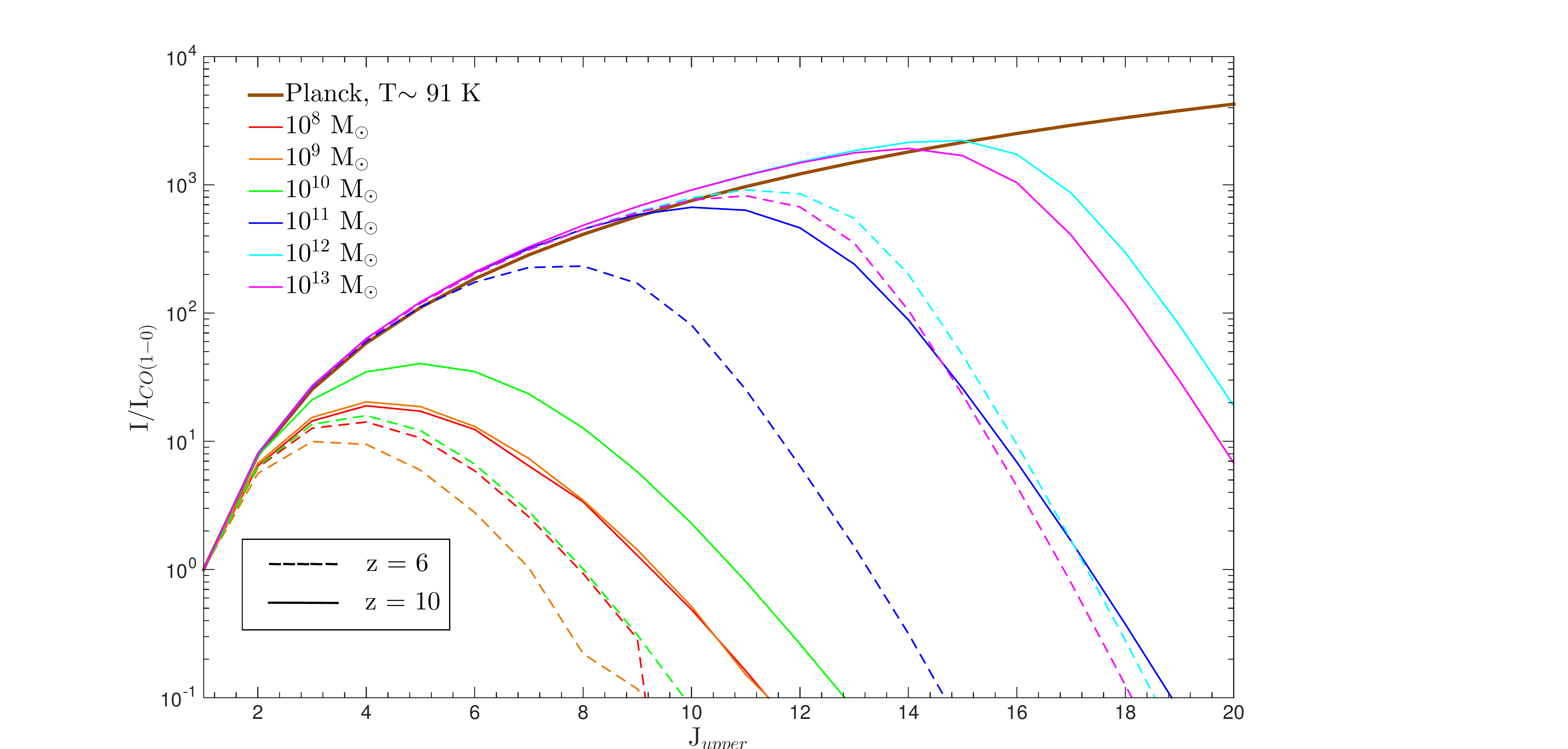}
\vspace{-.5cm}\caption{CO SLEDs generated by a halo at redshift $z$ = 6 (dashed curves) and $z$ = 10 (solid curves) with halo mass ranging from 10$^8$ to 10$^{13}$ M$_\odot$, neglecting the effects of photodissociation. If CO photodissociation is taken into account, the red and orange curves, corresponding to line emission from halos with $M <$ 10$^{10}$ M$_\odot$, would disappear while the other curves would remain unchanged. The brown curve denotes the expected SLED in the optically-thick case when the population levels are in local thermal equilibrium (LTE); in this limiting case, the intensity at a given level follows the Planck function, $B(\nu,T)d\nu = \frac{2h\nu^3}{c^2}\frac{1}{e^{h\nu/kT}-1}d\nu$, where $\langle T \rangle \sim$ 91 K.}
\end{minipage}
\end{figure*}

The curves plotted in figure 5 demonstrate this trend in the case where a UV duty cycle of unity is assumed. In this scenario, the SFR and SFR surface density range from 10$^{-4}$-1000 M$_\odot$\,yr$^{-1}$ and 10$^{-2}$-20 M$_\odot$\,yr$^{-1}$\,kpc$^{-2}$ respectively, when the halo mass varies from 10$^8$ to 10$^{13}$ M$_\odot$ (see figure 1). While the corresponding kinetic temperature remains relatively constant for this $\Sigma_{SFR}$ range ($T_{kin} \sim$ 90 K), the H$_2$ number density varies significantly, 10$^2$ cm$^{-3}$ $\lesssim n_{H_2} \lesssim 3\times10^5$ cm$^{-3}$. The variance in the shape of the CO SED computed for $f_{UV}$ = 1 reflects this range of physical conditions parameterized by the halo mass at $z$ = 10; the SEDs produced by low-mass halos, $M <$ 10$^{10}$ M$_\odot$, peak at $J \simeq$ 4 before turning over and plummeting (neglecting CO photodissociation). In these low density clouds, with correspondingly small velocity gradients ($dv/dr \sim$ 0.5 km\,s$^{-1}$\,pc$^{-1}$) and CO abundances ($\chi_{CO}\sim$ 5$\times$10$^{-6}$), only the low-lying transitions such as CO $J$ = 1\,$\rightarrow$\,0 are optically thick. In contrast, in more extreme star-forming galaxies, i.e. $M >$ 10$^{10}$ M$_\odot$, the H$_2$ number densities reach $\sim$ 10$^5$ cm$^{-3}$, the typical critical density value for high-$J$ CO emission.
Consequently, the high-$J$ lines grow optically thick and the line ratios approach thermalization for transitions as high as $J \sim$ 13 in these high-mass halos.

\FloatBarrier
\begin{figure*}[h!]
\begin{minipage}{1\linewidth}
\vspace{-1cm}
\hspace{-1.5cm}\includegraphics[width=275pt,height=300pt]{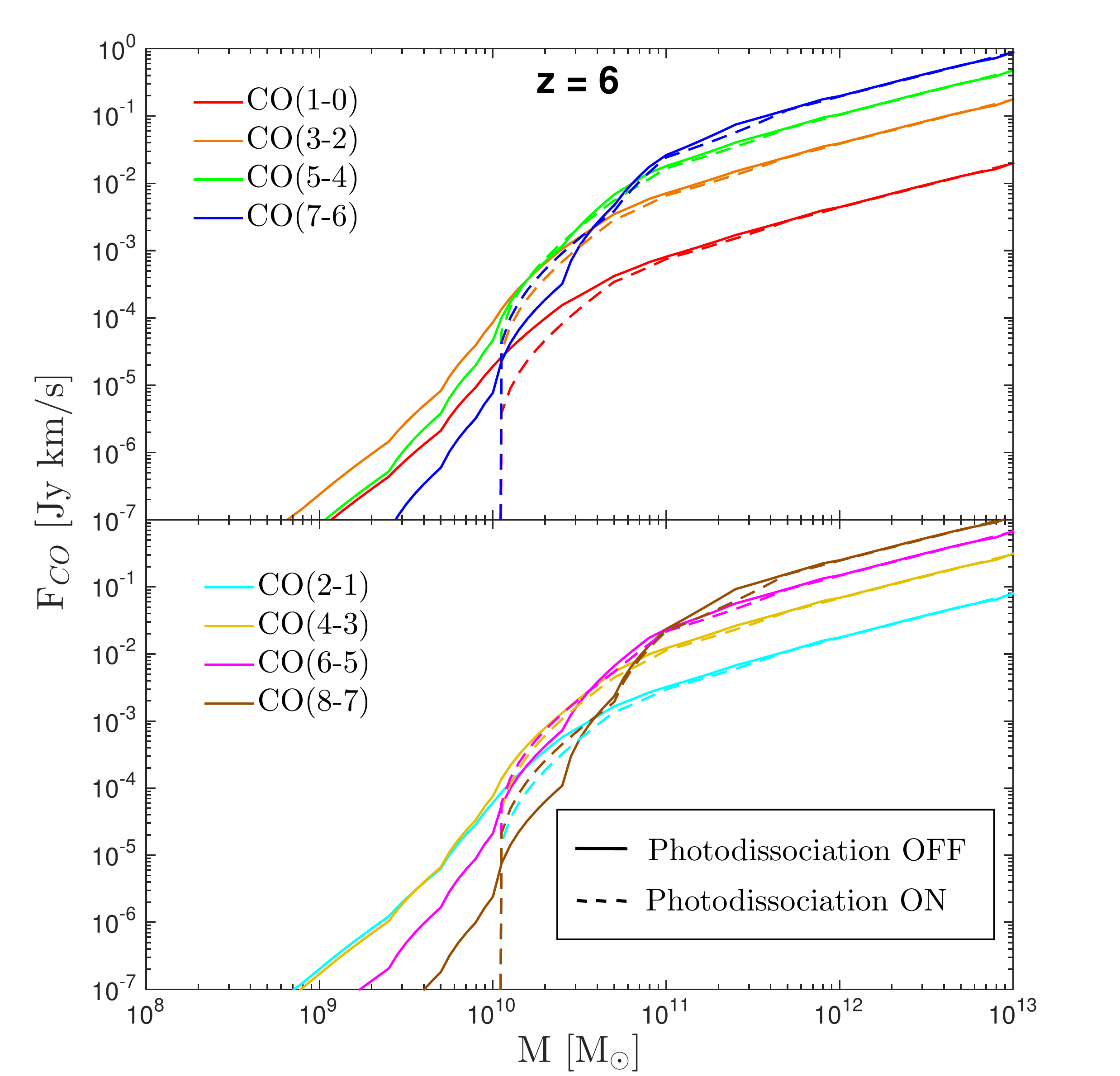}
\hspace{-.5cm}\includegraphics[width=275pt,height=300pt]{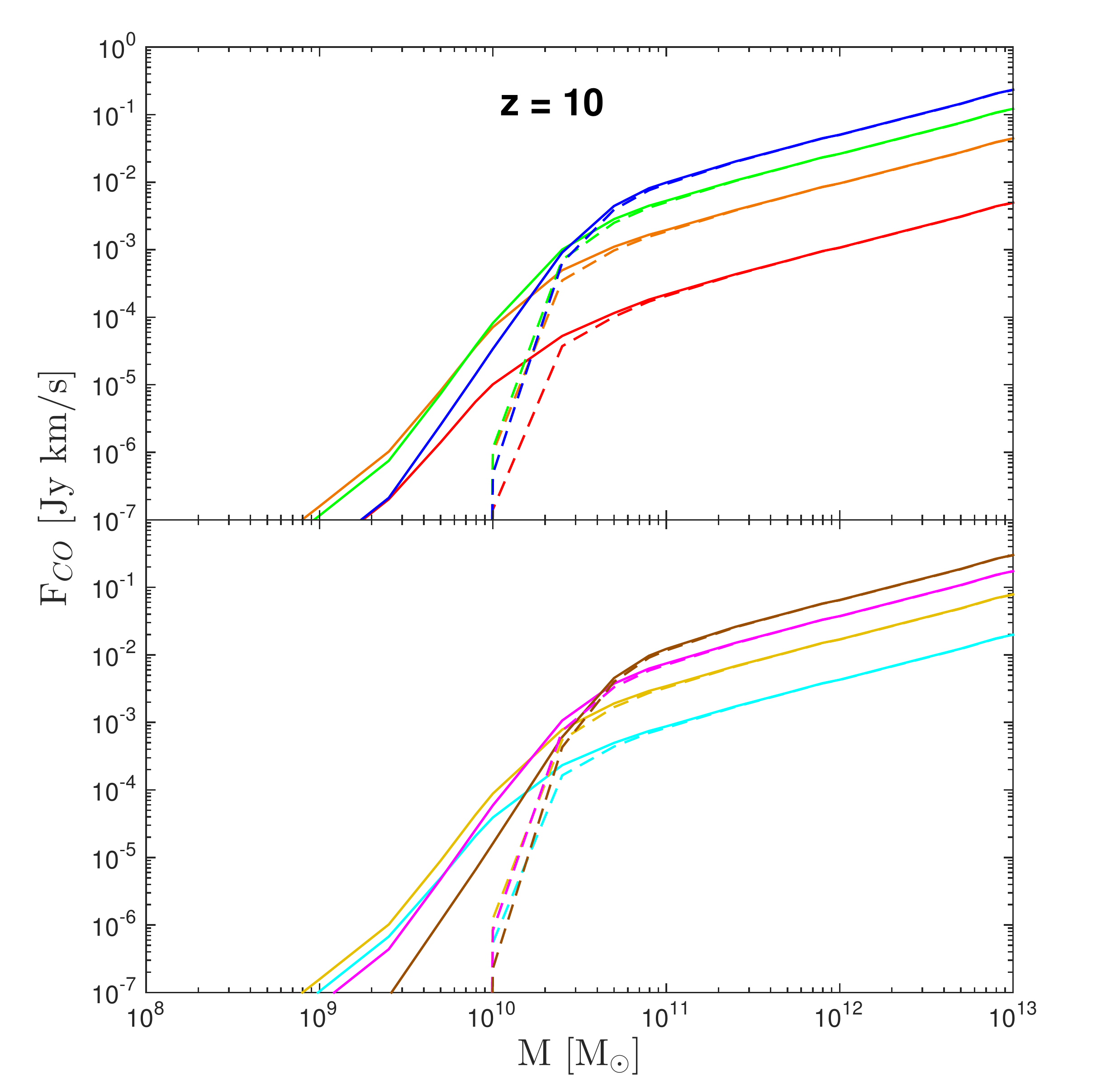}
\hspace{-1cm}\vspace{-.5cm}\caption{The CO rotational line flux as a function of halo mass at $z$ = 6 (left panel) and $z$ = 10 (right panel) for $J_{upper}$ = 1-8. Solid and dashed curves denote results obtained by neglecting (eq. (2.30)) and including (eq. (2.32)) the effects of CO photodissociation.}
\end{minipage}
\end{figure*}

\section{Results}

\subsection{Predicted CO Fluxes and Spatially Averaged CO Brightness Temperature}

Given our LVG model of the full CO SLED, we can now predict the CO line fluxes generated by a set of molecular clouds residing in a host halo of mass $M$ at redshift $z$, as well as compute the spatially averaged brightness temperature of any CO rotational transition. The former is obtained by converting LVG-derived CO luminosities (eq. (2.8)) into observed, velocity-integrated fluxes, $F_{CO}$, using typical observer units,
\begin{equation}
\frac{L_{CO}}{L_\odot} = 1.040\times10^{-3}\left(\frac{D_L}{\text{Mpc}}\right)^2\frac{\nu_{obs}}{\text{GHz}}\,\,\frac{F_{CO}}{\text{Jy km s$^{-1}$}}
\end{equation}
where $D_L$ is the luminosity distance. The results are shown in figure 6 for a range of CO rotational lines emitted as a function of host halo mass at $z$ = 6 (left panel) and $z$ = 10 (right panel). The dashed curves represent the fluxes obtained by including the effects of CO photodissociation in our computations; as expected, the change in flux when accounting for this phenomenon is most pronounced at the low-mass end where $F_{CO}$ drops to zero once the CO is fully photodissociated. 
At the higher-mass end, we predict that a $\sim$10$^{11}$ M$_\odot$ halo at redshift $z$ = 6 will emit strongest in the J =  7 $\rightarrow$ 6 transition with a flux of $\sim$ 25 mJy while a halo with the same mass at $z$ = 10 will emit strongest in a higher energy state, J =  10 $\rightarrow$ 9, but with nearly half the flux, $\sim$ 14 mJy.

\FloatBarrier
\begin{figure*}[h!]
\begin{minipage}{1\linewidth}
\vspace{-1cm}
\hspace{-2cm}\includegraphics[width=530pt,height=270pt]{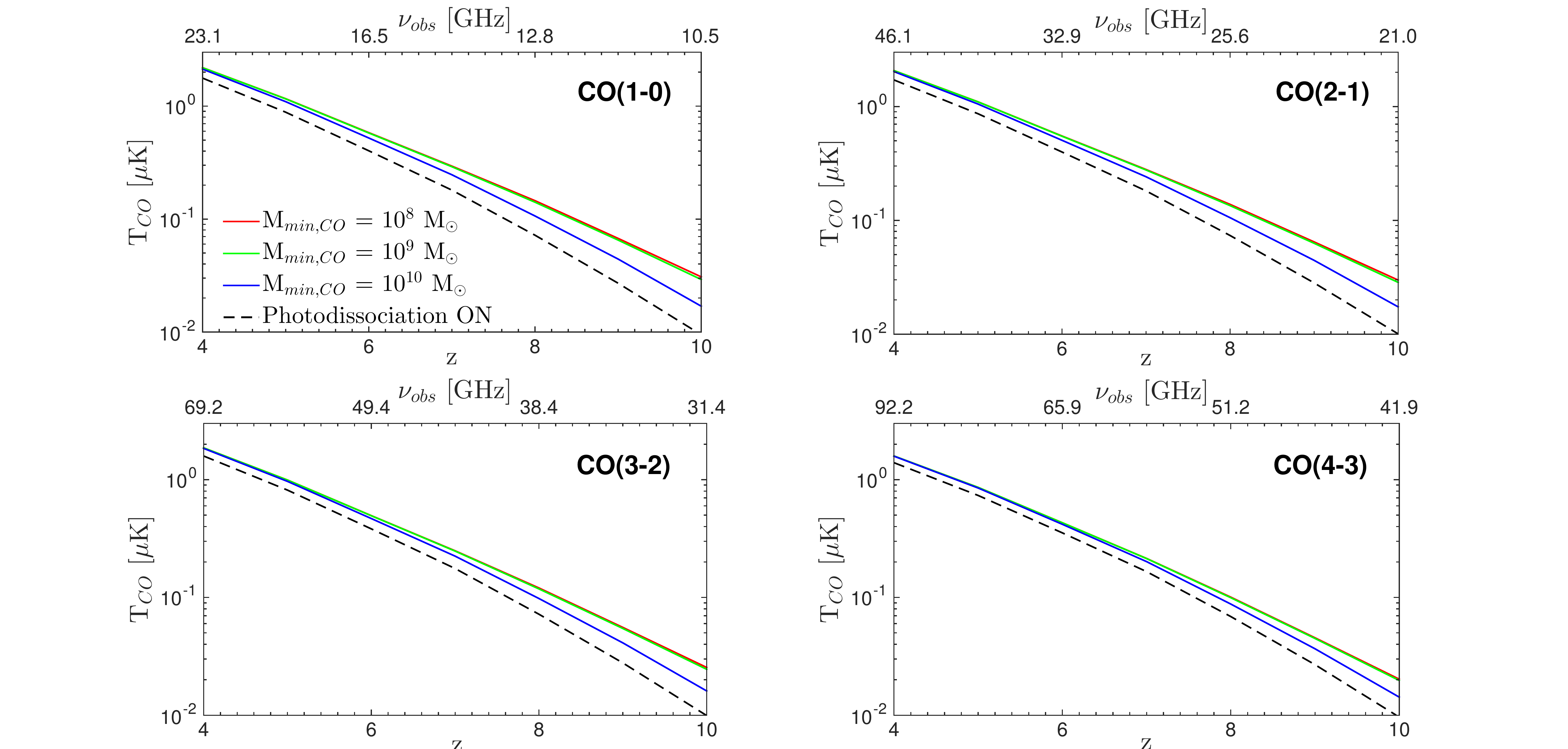}\\
\hspace*{-2cm}\includegraphics[width=530pt,height=270pt]{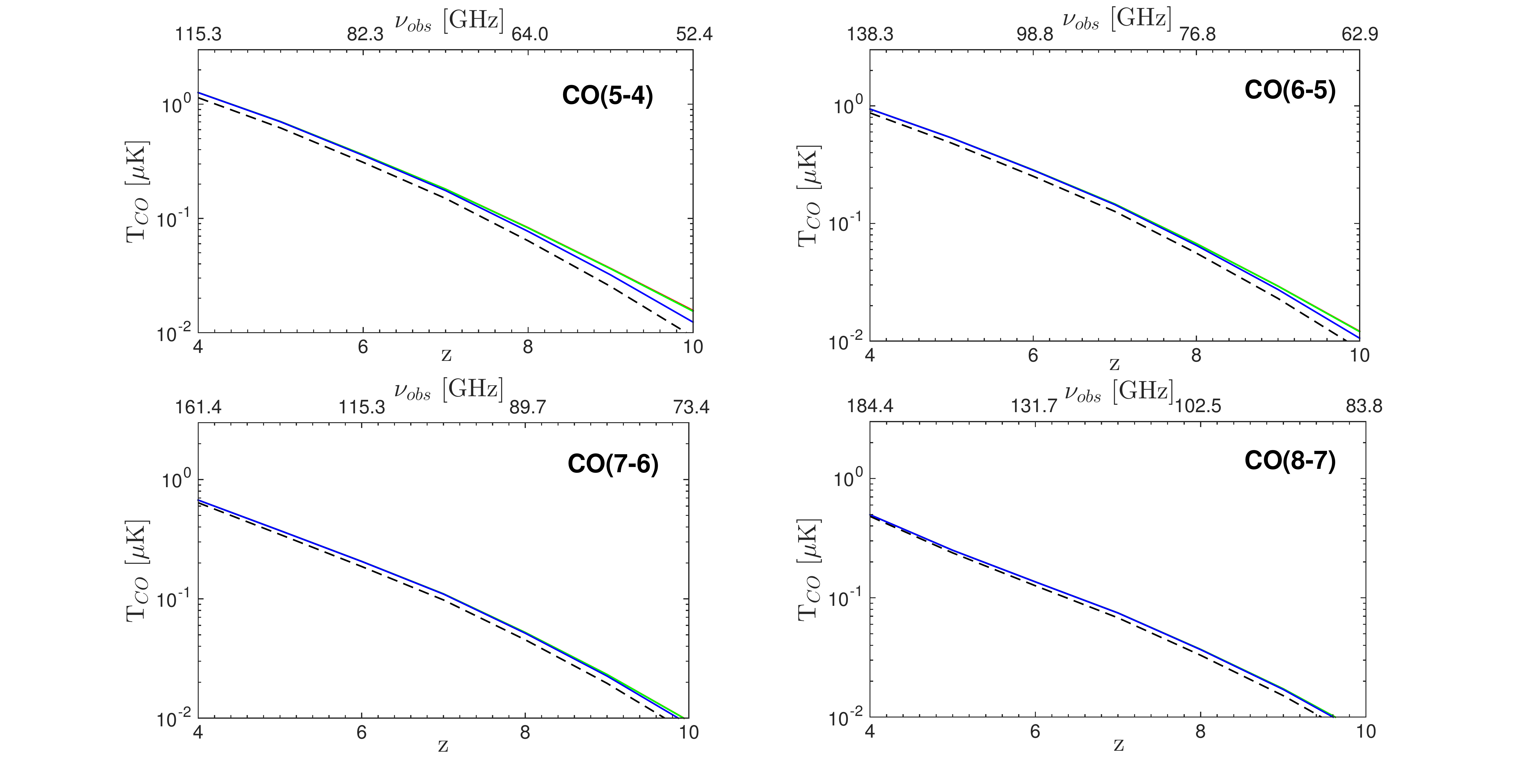}
\hspace{-1cm}\vspace{-1cm}\caption{Volume-averaged CO brightness temperature as a function of redshift for a minimum host halo mass of CO luminous galaxies of M$_{min,CO}$ = 10$^8$ (red), 10$^9$ (green), and 10$^{10}$ M$_\odot$ (blue curves), neglecting the effects of CO photodissociation. The dashed curves denote the signals obtained when photodissociation is taken into account (M$_{min,CO}$ = 10$^{10}$). Each panel shows $\langle T_{CO} \rangle$ for a different line in the CO rotational ladder with 1 $\leq J_{upper} \leq$ 8.}
\end{minipage}
\end{figure*}

In order to determine the spatially-averaged CO brightness temperature emitted by halos across the mass range, we compute the following integral,
\begin{equation}
\langle T_{CO}(\nu_J) \rangle = \frac{c^3}{8\pi}\frac{(1+z_J)^2}{k_B \nu_J^3 H(z_J)}f_{duty}\int_{M_{min,CO}}^\infty\hspace{-0.7cm} dM\,\,\frac{dn}{dM}(M,z_J)L(M,z_J) 
\end{equation}
where $\nu_J$ is the rest-frame frequency of the specified line. This equation is parameterized by $M_{min,CO}$, the minimum host halo mass for CO luminous halos, and $f_{duty}$, the duty cycle for CO activity. Since CO lines are excited by starburst activity, we generally expect that the duty cycle for CO luminous activity is comparable to the starburst duty cycle. We therefore assume $f_{duty} = f_{UV}$ = 1 in our fiducial models. Furthermore, in computing the volume-averaged CO brightness temperature, we vary $M_{min,CO}$ widely between $M_{min,CO}$ = 10$^8$, 10$^9$, and 10$^{10}$ M$_\odot$ to illustrate the sensitivity of the results to this parameter. In our model, these halos host CO luminous galaxies with SFRs of $\sim$ 10$^{-4}$, 3$\times$10$^{-3}$, and 0.1 M$_\odot$\,yr$^{-1}$ (left panel of figure 1).

We plot the results in figure 7, where the mean brightness temperature is shown as a function of redshift for different CO rotational lines. As expected, we find that $\langle T_{CO} \rangle$ is a

\noindent steeply declining function of $z$, a direct consequence of the decreasing number of host halos (per volume) at these high redshifts predicted by the Sheth-Tormen halo mass function~\cite{1999MNRAS.308..119S}. When CO photodissociation is neglected (solid curves), the redshift evolution of $\langle T_{CO} \rangle$ for the low-$J$ CO lines steepens at the high-$z$ end when larger values are assumed for $M_{min,CO}$.  

While halos with masses $M >$ 10$^{10}$ M$_\odot$ emit the most flux at all redshifts (figure 6), the flux emitted by halos with $M \sim$ 10$^9$-10$^{10}$ M$_\odot$ makes a non-negligible contribution to the total CO brightness temperature at higher redshifts. An illustration of this is presented in figure 8 which plots $dT/d\ln{M}$, the contribution to the mean brightness temperature per logarithmic halo mass for different CO lines as a function of halo mass at $z$ = 6 (dashed curves) and $z$ = 10 (solid curves). It is clear that at $z$ = 10, although the primary component of the CO signal originates from halos with masses in the range $M \sim$ 10$^{10}$-10$^{11}$ M$_\odot$, the CO emission from $ \sim 10^9$ M$_\odot$ halos makes up nearly 15\% of the total signal for the low-J lines. Thus, raising the minimum host halo mass for CO luminous halos from 10$^8$ M$_\odot$ to 10$^{10}$ M$_\odot$ excludes a population of CO-emitting sources and reduces the amplitude of $\langle T_{CO} \rangle$ for the low-J lines by a factor of $\sim$ 2. The effects of raising $M_{min,CO}$ vanish in the higher energy states since these low-mass halos emit less than 1\% of the total signal in these high-J lines; hence, the red, green, and blue solid curves, denoting the mean brightness temperature for models where $M_{min,CO}$ = 10$^8$, 10$^9$, and 10$^{10}$ M$_\odot$, respectively, are barely distinguishable from one another for the higher energy rotational transitions. 

In the case where CO photodissociation is taken into account, we found that the CO becomes fully photodissociated in halos with $M <$ 10$^{10}$ M$_\odot$ (figure 4); since these low-mass halos do not emit CO flux in this model, the minimum host halo mass of a CO luminous galaxy is effectively set to $M_{min,CO}$ = 10$^{10}$ M$_\odot$. Accounting for CO destruction in molecular clouds results in an overall reduction in the amplitude of $\langle T_{CO} \rangle$, with the low- and high-J lines weakening by $\sim$ 2-20\% and 10-45\%, respectively, over the range of redshifts shown in figure 7 (dashed curves).

\FloatBarrier
\begin{figure*}[t!]
\begin{minipage}{1\linewidth}
\hspace{2cm}\includegraphics[width=350pt,height=270pt]{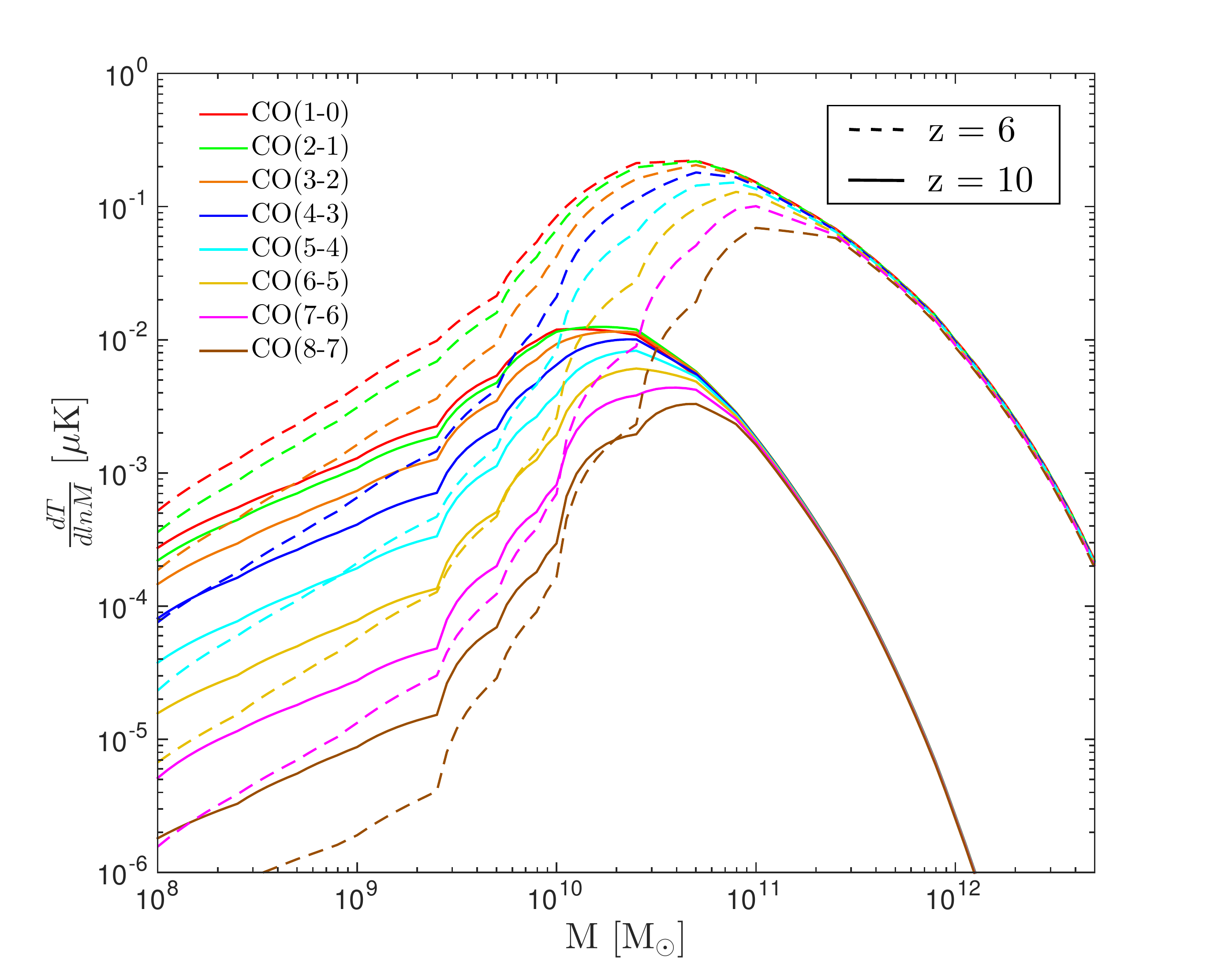}
\vspace{-.5cm}\caption{Contribution to the mean brightness temperature per logarithmic mass of different CO lines at redshift $z$ = 6 (dashed) and $z$ = 10 (solid), in the case where the effects of CO photodissociation are neglected.}
\end{minipage}
\end{figure*}

The spatially averaged brightness temperature of the CO J = 1$\rightarrow$\,0 line predicted by our model assuming $M_{min,CO}$ = 10$^8$ M$_\odot$  ranges from $\sim$ 0.6 $\mu$K at $z$ = 6 to $\sim$ 0.03 $\mu$K at $z$ = 10 when CO photodissociation is neglected. The strength of the CO signal dwindles for higher-J lines, with $\langle T_{CO}(z=6) \rangle \sim$ 0.3 $\mu$K and $T_{CO}(z=10) \rangle \sim$ 0.1 $\mu$K for the CO J = 6$\rightarrow$\,5 transition.  
``Turning on" photodissociation does not significantly effect the mean brightness temperature of the high-J lines, but it does reduce the CO(1-0) signal to $\langle T_{CO} \rangle$ $\sim$ 0.4 and $\sim$ 0.01 $\mu$K at $z$ = 6 and 10, respectively .

Without delving into any particular instrumental design, we briefly consider the plausibility of detecting such signals. We use the radiometer equation for the signal-to-noise ratio, $S/N = (T_{CO}/T_{sys})\sqrt{\Delta\nu \,t_{int}}$ where $T_{sys} \sim$ 20 K is the system temperature of the detector, $\Delta\nu$ is the observed bandwidth, and $t_{int}$ is the integration time, which we take to be 1000 hours. We find that at $z$ = 6, the predicted brightness temperatures of the 4 lowest-lying CO transitions (neglecting CO photodissociation) can be detected at 5$\sigma$ confidence with a bandwidth of $\Delta\nu\sim$ 10 GHz, while a bandwidth of $\Delta\nu \sim$ 100 GHz is required for detection with 1$\sigma$ confidence at $z$ = 10. Higher J lines in this model will require even longer integration times to achieve the desired brightness sensitivity.

\subsection{CO Power Spectrum}
Given the existence of brighter foreground sources of emission at the relevant frequencies, the mean redshifted CO signal will be difficult, if not impossible, to directly observe. We therefore consider spatial fluctuations in the surface brightness and compute the CO power spectrum predicted by different variations of our model. In contrast to the spectrally smooth foreground sources, the CO signal is expected to have structure in frequency space which can be used to isolate spatial fluctuations in its brightness temperature. Since the power spectrum captures the underlying matter distribution and structure, a map of the CO brightness temperature fluctuations at $z \geq$ 6 can be used to probe the spatial distribution of star-forming galaxies during the EoR. 

Following the formalism in~\cite{2011ApJ...728L..46G} and~\cite{2011ApJ...741...70L}, the three-dimensional power spectrum of the CO brightness temperature fluctuations is expected to take the form
\begin{equation}
P_{CO}(k,z) = \langle T_{CO} \rangle(z) \left[ b_{CO}(z)^2 P_{lin}(k,z)+P_{shot}(z)\right] \,\,.
\end{equation}
Since CO is emitted from within halos, the first term in this expression represents spatial variations due to correlations with the underlying dark matter density field  where $P_{lin}$ is the linear theory density power spectrum and the bias $b_{CO}$ is given by
\begin{equation}
b_{CO}(z) = \frac{\int_{M_{min,CO}}^\infty\hspace{-.5cm} dM \frac{dn}{dM}L_{CO}(M,z)b(M,z)}{\int_{M_{min,CO}}^\infty \hspace{-.5cm}dM \frac{dn}{dM}L_{CO}(M,z)}
\end{equation}
where $b(M,z) = 1 + (\nu^2(M,z)-1)/\delta_c$, $\nu(M,z)=\delta_c/\sigma(M,z)$, and $\sigma(M,z)$ is the RMS density fluctuation in a spherical region containing mass $M$~\cite{2002MNRAS.336..112M}. The second term of eq. (3.3), the shot noise contribution due to Poisson fluctuations in the number of halos on the sky, can be expressed as
\begin{equation}
P_{shot}(z)=\frac{1}{f_{duty}}\frac{\int_{M_{min,CO}}^\infty\hspace{-.5cm}dM \frac{dn}{dM}L_{CO}(M,z)^2}{\left(\int_{M_{min,CO}}^\infty\hspace{-.5cm}dM \frac{dn}{dM}L_{CO}(M,z)\right)^2}
\end{equation}
where, in all of our calculations, we adopt the Sheth-Tormen halo mass function for $dn/dM$.

\FloatBarrier
\begin{figure*}[t!]
\begin{minipage}{1\linewidth}
\vspace{-2cm}
\hspace{2cm}\includegraphics[width=300pt,height=200pt]{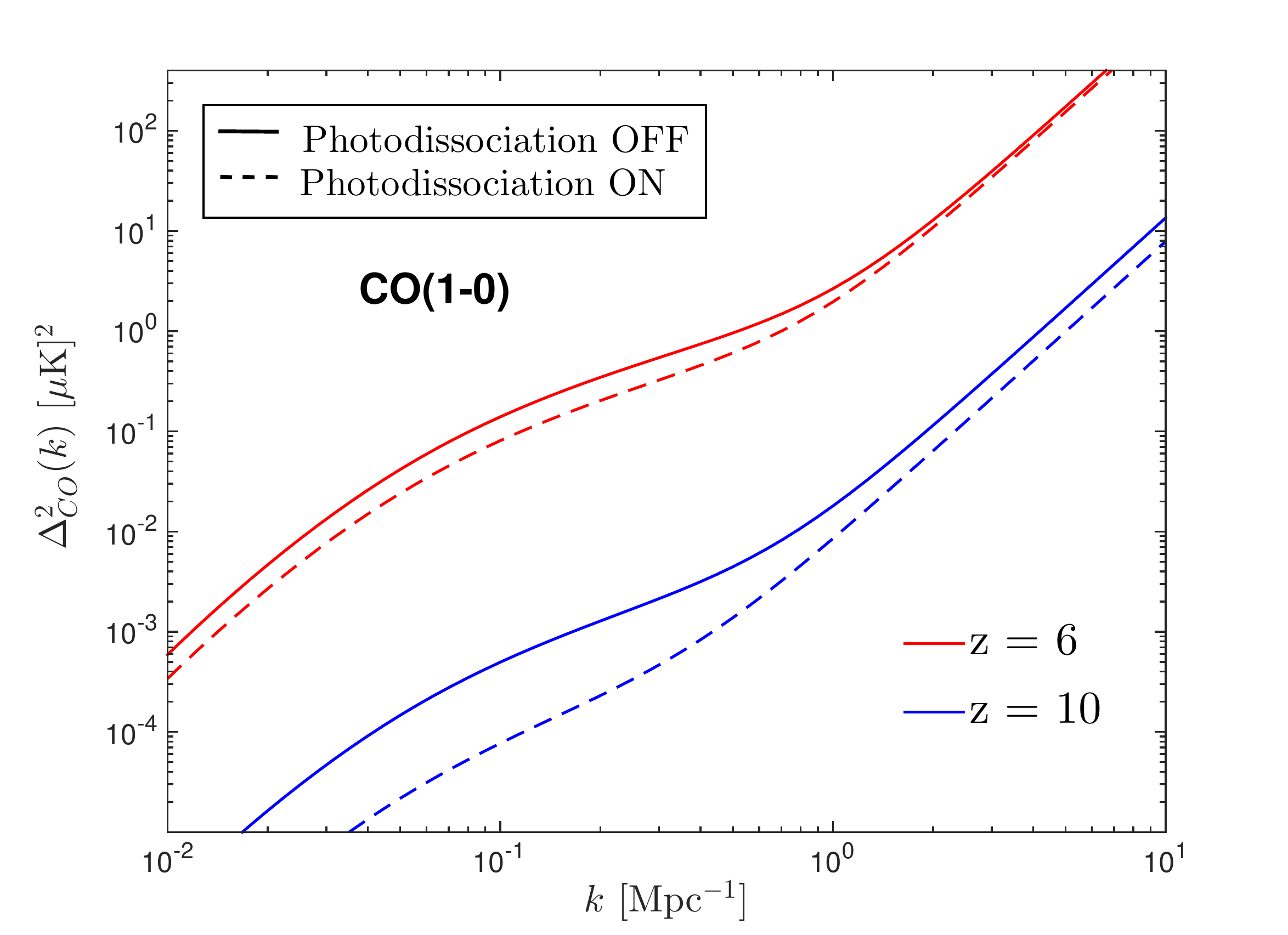}
\end{minipage}
\begin{minipage}{1\linewidth}
\hspace{-3cm}\includegraphics[width=600pt,height=300pt]{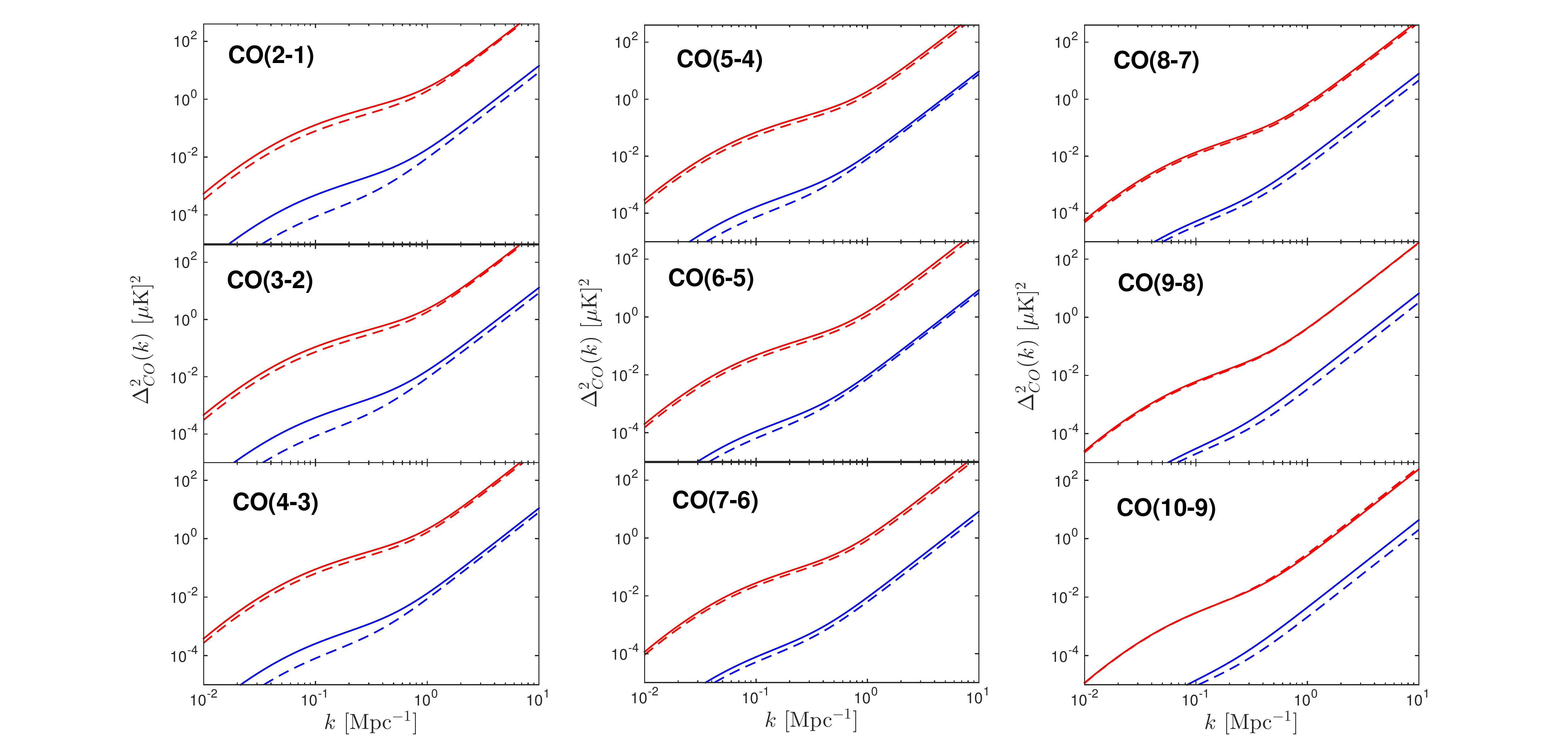}
\vspace{-0.5cm}\caption{Auto power spectrum of CO brightness temperature fluctuations for lines all the way up to $J_{upper}$ = 10, at $z$ = 6 (red) and $z$ = 10 (blue). In each panel, the solid curves represent results for the models in which CO photodissociation was neglected and a minimum host halo mass of $M_{CO,min}$ = 10$^8$ M$_\odot$ was assumed; dashed curves denote results for the case where the effects of CO photodissociation were included, effectively setting the minimum host halo for CO luminous galaxies to $M_{CO,min}$ = 10$^{10}$ M$_\odot$.}
\end{minipage}
\end{figure*}

The power spectra of the CO rotational lines from $J_{upper}$ = 1 to $J_{upper}$ = 10 are plotted in figure 9 for models which include (dashed) and exclude (solid) the effects of CO photodissociation. The y-axis shows $\Delta^2_{CO}(k,z) = k^3 P_{CO}(k,z)/(2\pi^2)$, the contribution to the variance of $\langle T_{CO} \rangle$ per logarithmic bin, in units of [$\mu$K$^2$]. 
In general, the fluctuations depend strongly on the wavenumber, with the overall shape of the predicted power spectra reflecting the form of eq. (32). On large scales, small $k$, the clustering term dominates and the fluctuations in CO brightness temperature mirror the underlying dark matter density field. Conversly, on small scales (large $k$), the shot noise takes over and the fluctuations reach $\Delta^2_{CO} \sim$ 300 - 1300 $\mu$K$^2$ on scales of $k$ = 10 Mpc$^{-1}$ at $z$ = 6, and $\Delta^2_{CO} \sim$ 4 - 14 $\mu$K$^2$ at $z$ = 10 (when photodissociation is neglected). The $J$ = 1$\rightarrow$\,0 transition is predicted to produce the strongest signal, with an amplitude that drops by a factor of  $\sim$5-50 across the wavenumber range $k \sim$ 10$^{-2}$-10 Mpc$^{-1}$ when the higher energy state, $J$ = 10$\rightarrow$\,9, is considered instead.

 The redshift evolution of $\Delta^2_{CO}$ is ultimately dictated by the behavior of $\langle T_{CO} \rangle(z)$; although $b_{CO}^2$ increases as the host halos become more clustered at higher redshifts, this effect is not enough to compensate the declining brightness temperature with $z$. The auto-correlation signal of all the CO lines therefore weakens as the redshift varies from $z$ = 6 to 10, dropping by $\sim$ 2 orders of magnitude over this redshift range. On the other hand, our results for $\Delta^2_{CO}$ depend very weakly on our choice of the minimum host halo mass for CO luminous galaxies; therefore, in the fiducial models where photodissociation is neglected, we only plot results for the case where $M_{min,CO}$ = 10$^8$ M$_\odot$, thereby including the contribution of the low-mass halos to the overall emission signal.
 
 The effects of including CO photodissociation are most prominent for the low-$J$ lines, weakening the signal by, at most, a factor of $\sim$ 6 at $z$ = 10 on large scales. At $k$ = 0.1 Mpc$^{-1}$, we find CO(1-0) brightness temperature fluctuations of amplitude $\Delta_{CO}^2 \sim$ 0.1 $\mu$K$^2$ at $z$ = 6, whether or not CO photodissociation is taken into account. At $z$ = 10,  $\Delta_{CO}^2(k=0.1$ Mpc$^{-1})$ $\sim$ 5$\times$10$^{-4}$ and 8$\times$10$^{-5}$ $\mu$K$^2$ for the $J$ = 1$\rightarrow$\,0 line in the models where photodissociation is turned ``off" (assuming fixed $M_{min,CO}$ = 10$^8$ M$_\odot$) and ``on", respectively.

\section{Discussion}
We have presented a new approach to estimating the mean CO emission signal from the epoch of reionization (EoR) that links the atomic physics of molecular emission lines to high-redshift observations of star-forming galaxies. This method is based on LVG modeling, a radiative transfer modeling technique that generates the full CO SLED for a specified set of characterizing parameters, namely, the kinetic temperature, number density, velocity gradient, CO abundance, and column density of the emitting source. We showed that these LVG parameters, which dictate both the shape and amplitude of the CO SLED, can be expressed in terms of the emitting galaxy's global star formation rate, $SFR$, and the star formation rate surface density, $\Sigma_{SFR}$. Employing the $SFR$-$M$ relation empirically derived for high-redshift galaxies, i.e. $z \geq$ 4, we can then ultimately express the LVG parameters, and thus, the specific intensity of any CO rotational line, as functions of the host halo mass $M$ and redshift $z$. 

Adopting a starburst duty cycle of $f_{UV}$ = 1, the average $SFR-M$ relation derived via abundance-matching for 4 $< z <$ 8 is characterized by a steeply declining slope at the low-mass end, where the star formation rate goes as $SFR \propto M^{1.6}$. With the SFR dropping to values below 0.1 M$_\odot$\,yr$^{-1}$ for $M <$ 10$^{10}$ M$_\odot$, the physical conditions in these low-mass halos are not ``extreme" enough to substantially excite the high-$J$ CO rotational states. The resulting CO SLEDs correspondingly peak around $J \simeq$ 4 before turning over, and the overall contribution of CO emission from this halo population grows negligible at lower redshifts. On the other hand, the H$_2$ number density and CO-to-H$_2$ abundance in halos with masses $M \geq$ 10$^{10}$ M$_\odot$ are large enough to keep the high-$J$ population levels thermalized; the CO SLEDs generated by these massive halos therefore have peaks shifited to $J \geq$ 10 and overall higher amplitudes, reflecting the excitation of the more energetic states of the molecule.  

We also consider the effects of CO photodissociation on the CO line intensities generated by our fiducial models. Assuming that the threshold for CO survival is set by a universal dust opacity, we adopt a first-order approximation of $N_{CO}$ which effectively reduces the CO column density with decreasing metallicity. In the limiting case where the metallicity drops so low that there is not enough dust to shield CO, the CO is considered fully dissociated and $N_{CO}$ is set to zero. We find that such conditions occur in halos with $ M \lesssim$ 10$^{10}$ M$_\odot$, causing the line intensities emitted by these low-mass halos to entirely disappear in models where CO photodissociation is accounted for. 

Given our LVG model of the full CO SLEDs, we can predict both $F_{CO}$, the CO line flux generated by a set of molecular clouds in a host halo of mass $M$ at redshift $z$, as well as $\langle T_{CO} \rangle$, the spatially averaged brightness temperature of any CO rotational transition. We find that the flux emitted in the $J$ = 1 $\rightarrow$ 0 transition by a halo at redshift $z$ = 6 with mass $M \sim$ 10$^{11}$ M$_\odot$ is $F_{CO(1-0)} \sim$ 0.8 mJy km/s. The higher rotational lines are expected to be even brighter, with a 10$^{11}$ M$_\odot$ halo at $z$ = 6 emitting a CO(6-5) flux of $\sim$ 23 mJy km/s. These fluxes drop by 25-30\% when considering a 10$^{11}$ M$_\odot$ halo emitting at $z$ = 10, with $F_{CO(1-0)} \sim$ 0.2 mJy km/s and $F_{CO(6-5)} \sim$ 7 mJy km/s. 

The CO line fluxes emitted by individual host halos of mass $M$ at redshift $z$ as estimated in this paper, are found to be generally higher than previous estimates obtained in the literature ~\cite{2009ApJ...698.1467O,2013MNRAS.435.2676M}. 
Building an analytic formalism within a paradigm where star formation is a function of gas supply and stellar feedback, the model presented in~\cite{2013MNRAS.435.2676M} provides radial distributions of SFR, $\Sigma_{gas}$, and $T$ which are then used to calculate the masses, sizes, and number of GMCs, and the corresponding total CO line luminosities emitted by these clouds.
Consequently, the CO(1-0) flux emitted by a 10$^{11}$ M$_\odot$ halo at $z$ = 6 as predicted by this model is $\sim$ 0.01-0.03 mJy km/s, an order of magnitude smaller than the flux estimates derived with our methodology. The flux in the $J$ = 6 $\rightarrow$ 5 transition is found to be highly sensitive to the inclusion of turbulent clumps within the GMCs, given that the high densities in such inhomogeneities can effect the thermalization of the higher-$J$ rotational lines; the resulting CO(6-5) flux in the models presented in~\cite{2013MNRAS.435.2676M} thus varies from 10 mJy km/s to 10$^{-3}$ mJy km/s when turbulent clumps are included and excluded, respectively. 

While these results are comparable to those derived using the semi-analytic methods of~\cite{2009ApJ...698.1467O}, they are consistently smaller than the values presented in this paper. The prolific number of parameters used to characterize the GMC properties in~\cite{2013MNRAS.435.2676M} make it difficult to ascertain the source of divergence between our results, which both use radiative transfer techniques. However, we suspect these differences can be traced back to the different prescriptions used to set the star formation rate surface density, $\Sigma_{SFR}$, and gas surface density, $\Sigma_{gas}$, the two ingredients which, once related to one another via the Kennicutt-Schmidt relation in our model, determine the shape and amplitude of the resulting CO SLED. 
Furthermore,~\cite{2013MNRAS.435.2676M} parameterizes the properties of MCs as a function of their position relative to the disk center, and after considering the level populations and optical depth of CO in each cloud directly, computes the resulting flux from the entire galaxy by counting up the number of clouds at each galactic radius. The molecular clouds in our model, on the other hand, are characterized by disk-averaged properties of the host halo disk; to derive the signal from a CO-emitting galaxy, we assume a large number of these identical homogeneous collapsing clouds and scale the line intensities with the disk-averaged CO column density. 
Based on the sensitivity limits quoted in~\cite{2013MNRAS.435.2676M}, our approach predicts that the $J_{upper}$ = 1 line in $z$ = 6 halos with mass $M \geq$ 10$^{12}$ M$_\odot$ will be observable by JVLA\footnote{http://www.vla.nrao.edu/} (Jansky Very Large Array)  after ten hours of observation; at this redshift, the CO(2-1) and CO(3-2) lines fluxes emitted by halos with $M >$ 10$^{11}$ M$_\odot$ are also expected to be observed by JVLA. Similarly, we expect ALMA to detect the CO(6-5) flux emitted by $z$ = 6 halos with mass $M \geq$ 5$\times$10$^{10}$ M$_\odot$ after ten hours of observation, and higher rotational lines $J_{upper} \geq$ 7 emitted by halos with $M \gtrsim$ 10$^{11}$ M$_\odot$.

To obtain an estimate of the spatially averaged brightness temperature of a given line at a particular redshift, we simply integrate $L_{CO}(M,z)$ over the range of halo masses that are expected to host CO-luminous galaxies. In the case where CO photodissociation is included, the minimum host halo mass of CO-emitting galaxies is set to $M_{CO,min} \sim$ 10$^{10}$ M$_\odot$ by the model itself, since the CO is found to be fully dissociated in halos with $M <$ 10$^{10}$ M$_\odot$. When CO photodissociation is ignored, varying  $M_{CO,min}$ from 10$^8$ to 10$^{10}$ M$_\odot$ reduces the low-$J$ line signals at high redshifts, where the CO emission from $\sim$ 10$^9$ M$_\odot$ halos make up a non-negligible percentage of the total emission.

In our fiducial model where CO photodissociation is neglected and $M_{min,CO}$ = 10$^8$ M$_\odot$, we predict a spatially averaged brightness temperature of $\langle T_{CO} \rangle \sim$ 0.5 $\mu$K at $z$ = 6 and 0.03 $\mu$K at $z$ = 10 for the low-$J$ CO rotational lines, with brightness temperature fluctuations of amplitude $\Delta^2_{CO} \sim$ 0.1 and 0.005 $\mu$K$^2$ respectively, at $k$ = 0.1 Mpc$^{-1}$. These CO emission signals are further reduced to $\langle T_{CO} \rangle \sim$ 0.4 and 0.01 $\mu$K at $z$ = 6 and 10, respectively, for the low-lying states when the effects of CO photodissociation are included in the calculations. 
(Note that, since CO lines are typically excited by starburst activity, the choice of $M_{min,CO}$ = 10$^8$ M$_\odot$ is favored by theoretical and numerical investigations which indicate that 10$^8$ M$_\odot$ is the minimum mass required for a halo to cool and form stars at these high redshifts~\cite{1996ApJ...464..523H,1997ApJ...474....1T,2010ApJ...714L.202T,2013fgu..book.....L}).
Our estimates of $\langle T_{CO} \rangle$ for the low-$J$ CO transitions are comparable to the values obtained in previous work by~\cite{2011ApJ...730L..30C},~\cite{2011ApJ...728L..46G},~\cite{2011ApJ...741...70L}, and~\cite{2013ApJ...768...15P}. Constructing a model based on the required cosmic star formation rate density to reionize the universe,~\cite{2011ApJ...730L..30C} obtains an order-of-magnitude estimate of  $\langle T_{CO} \rangle(z=8) \sim$ 1 $\mu$K for the $J$ = 1$\rightarrow$\,0 and $J$ = 2$\rightarrow$\,1 transitions.~\cite{2011ApJ...728L..46G} arrives at a slightly smaller estimate of the CO(1-0) brightness temperature, $\langle T_{CO} \rangle \sim$ 0.5 $\mu$K for $z$ = 6 and 0.1 $\mu$K for $z$ = 10, by using the Millennium numerical simulation results of~\cite{2009ApJ...698.1467O} to model the CO emission from high-redshift galaxies.~\cite{2011ApJ...741...70L} assumes a linear $SFR-M$ relation and a set of low-$z$ empirical scaling relations between a galaxy's SFR, $L_{FIR}$, and $L_{CO(1-0)}$ to estimate $\langle T_{CO} \rangle$ for these low-$J$ lines; they find a mean brightness temperature of $\sim$ 2 $\mu$K at $z \sim$ 6 and 0.5 $\mu$K at $z \sim$ 10 with fluctuations $\Delta^2_{CO} \sim$ 0.2 and 0.02  $\mu$K$^2$, respectively, on scales of $k \sim$ 0.1 Mpc$^{-1}$. In a more recent paper,~\cite{2013ApJ...768...15P} follows the approach taken in~\cite{2011ApJ...741...70L} with a few adjustments to the $SFR-M$ prescription to arrive at brightness temperatures of $\langle T_{CO} \rangle \sim$ 0.7 and 0.2 $\mu$K at redshifts $z$ = 6 and 10 respectively, assuming $M_{min,CO}$ = 10$^9$ M$_\odot$ for the CO(1-0) and CO(2-1) lines. 

While these previous works are limited almost exclusively to predicting the signals of the $^{12}$CO J=1$\rightarrow$0 and 2$\rightarrow$1 transition lines, our LVG-based approach generates the full CO SLED and thus allows us to compute the signal strength of the higher-J energy states as well. For example, we predict a CO(10-9) brightness temperature of $\langle T_{CO} \rangle \sim$ 0.05 $\mu$K at $z$ = 6 and 0.003 $\mu$K at $z$ = 10 with $\Delta^2_{CO}(k = 0.1) \sim$ 0.003 and 10$^{-5}$ $\mu$K$^2$ respectively. We look forward to future experiments, such as the Carbon MonOxide Mapping Array (COMA)\footnote{http://www.stanford.edu/group/church_group/cgi-bin/wordpress/?page_id=515} currently under development, which promise to provide spectral-spatial intensity mapping of CO at the high redshifts characterizing the epoch of reionization.

\acknowledgments
We thank Reinhard Genzel for helpful discussions and suggestions.This work was supported by the Raymond and Beverly Sackler Tel Aviv University-Harvard/ITC Astronomy Program. A.L. acknowledges support from the Sackler Professorship by Special Appointment at Tel Aviv University. This work was also supported in part by a PBC Israel Science Foundation I-CORE Program grant 1829/12, and in part by NSF grant AST-1312034. This material is based upon work supported by the National Science Foundation Graduate Research Fellowship under Grant No. DGE1144152. Any opinion, findings, and conclusions or recommendations expressed in this material are those of the authors and do not
necessarily reflect the views of the National Science Foundation.

\bibliography{mybib}{}
\bibliographystyle{JHEP}

\end{document}